%% file: journal_v9.tex
\documentclass[onecolumn, draftcls, a4paper]{IEEEtran}

\pdfoutput=1 
\usepackage{color}
\usepackage{amsmath}
\usepackage{amssymb,amsfonts}
\usepackage{graphicx,subfigure}
\usepackage{cite,url,comment}
\usepackage{multirow}
\usepackage{rotating}
\usepackage[table]{xcolor}

\newtheorem{lemma}{Lemma}[section]
\newtheorem{theorem}{Theorem}[section]

\newcommand{\EE}{\mathbb{E}} % average operator
\newcommand{\RR}{\mathbb{R}} % real set
\newcommand{\ee}{{\rm e}}
\newcommand{\dd}{{\rm\,d}} % differential (for integrals)

\newcommand{\Lc}{{\cal L}}

\newcommand{\Pc}{{\cal P}}

\def\ben{\begin{enumerate}}
\def\beq{\begin{equation}}
\def\beqa{\begin{eqnarray}}
\def\bit{\begin{itemize}}
\def\een{\end{enumerate}}
\def\eeq{\end{equation}}
\def\eeqa{\end{eqnarray}}
\def\eit{\end{itemize}}

\def\non{\nonumber\\}

\usepackage{array}
\usepackage{amsfonts}

\newcommand{\insertfig}[4]{
\begin{figure}[t]
\centerline{\includegraphics[width=#1\columnwidth]{#2}}
\vspace{-3mm}\caption{#3}\label{#4}\vspace{-5mm}\end{figure}}

\newcommand{\ls}[1]
   {\dimen0=\fontdimen6\the\font
    \lineskip=#1\dimen0
    \advance\lineskip.5\fontdimen5\the\font
    \advance\lineskip-\dimen0
    \lineskiplimit=.9\lineskip
    \baselineskip=\lineskip
    \advance\baselineskip\dimen0
    \normallineskip\lineskip
    \normallineskiplimit\lineskiplimit
    \normalbaselineskip\baselineskip
    \ignorespaces
   }

\def\ben{\begin{enumerate}}
\def\beq{\begin{equation}}
\def\beqa{\begin{eqnarray}}
\def\bit{\begin{itemize}}
\def\een{\end{enumerate}}
\def\eeq{\end{equation}}
\def\eeqa{\end{eqnarray}}
\def\eit{\end{itemize}}

\def\non{\nonumber\\}

\DeclareMathAlphabet{\mathsfbf}{OT1}{cmss}{sbc}{n}

\newcommand{\Prmax}{p^{\rm max}}
\newcommand{\Prav}{\bar{p}}

\newcommand{\Psmax}{P^{\rm max}}
\newcommand{\Pcsmax}{\Pc^{\rm max}}

\newcommand{\Psav}{\bar{P}}
\newcommand{\Pcsav}{\bar{\Pc}}
\begin{document}

%\ls{2}

\title{Optimal Power Allocation Strategies in\\ Full-duplex Relay Networks\vspace*{-6pt}}
\author{Alessandro Nordio,~\IEEEmembership{Member,~IEEE,}
Carla Fabiana~Chiasserini,~\IEEEmembership{Senior~Member,~IEEE,}
Emanuele~Viterbo,~\IEEEmembership{Fellow,~IEEE}% <-this % stops a space
\thanks{A. Nordio is with CNR-IEIIT, Italy. C. F. Chiasserini is with
  Politecnico di Torino and a Research Associate at  CNR-IEIIT,
  Italy. E. Viterbo is with Monash University, Australia.}%
}

\date{}

\maketitle

\begin{abstract}

% CASO AF: da fare letter separata su questo

% NO BUFFER AT RELAY

%- $t_1/T$  $t_2/T$ probabilities of the PMF

%- linkare bene il risultato delle delta con i 2 fractions of time e
%livelli di potenza 

%- add disegno di fianco alla tabella: vedi disegno.

  In this work, we consider a dual-hop, decode-and-forward network
  where the relay can operate in full-duplex (FD) or half-duplex (HD)
  mode. We model the residual self-interference as an additive
  Gaussian noise with variance proportional to the relay transmit
  power, and we assume a Gaussian input distribution at the
  source. Unlike previous work, we assume that the source is only
  aware of the transmit power distribution adopted by the relay over a
  given time horizon, but not of the symbols that the relay is
  currently transmitting. This assumption better reflects the
  practical situation where the relay node also forwards signaling
  traffic, or data originated by other sources.  Under these
  conditions, we identify the optimal power allocation strategy at the
  source and relay, which in some cases coincides with the half duplex
  transmission mode. In particular, we prove that such strategy
  implies either FD transmissions over an entire frame, or FD/HD
  transmissions over a variable fraction of the frame.  We determine
  the optimal transmit power level at the source and relay for each
  frame, or fraction thereof. We compare the performance of the
  proposed scheme against reference FD and HD techniques, which assume
  that the source is aware of the symbols instantaneously transmitted
  by the relay. Our results highlight that our scheme closely
  approaches or outperforms such reference strategies.
\end{abstract}

\section{Introduction}
Multi-hop relay communications are a key technology for next
generation wireless networks, as they can extend radio access in case
of coverage holes or users at the cell edge, as well as increase the
potentialities of device-to-device data transfers.  The dual-hop relay
channel, in particular, has been widely investigated under different
cooperative schemes, namely, decode-and-forward (DF),
compress-and-forward (CF) and amplify-and-forward (AF) 
\cite{ElGamal06,Zafar14,Tuninetti14,Tuninetti16,Wang17}.  Most of this
body of work has assumed the relay node to operate in half-duplex (HD)
mode. Specifically, results on the capacity of the HD dual-hop relay
channel have appeared in \cite{Kramer04,Zlatanov15}, where it was
shown that the network capacity is achieved by a discrete input when
no direct link between the source and the destination exists.

More recently, a number of studies
\cite{Riihonen11-1,Day12,Kang14,Alouini2016,Zlatanov2017} have
addressed the case where the relay operates in full-duplex (FD) mode,
i.e., it can transmit and receive simultaneously on the same frequency
band. Indeed, advances in self-interference suppression in FD systems
have made such a technology very attractive for relay networks.  The
capacity of Gaussian two-hop FD relay channels has been characterized
in \cite{Cover79}, under the assumption that the residual
self-interference can be neglected. The more realistic case where
residual self-interference (\cite{Riihonen11-2,Korpi15}) is taken into
account, has been instead addressed in
\cite{Riihonen11-1,Day12,Kang14,Alouini2016,Zlatanov2017}. In these
works, the signal looping back from the relay output to its input is
modeled as an additive Gaussian noise with variance proportional to
the relay transmit power. In particular, \cite{Riihonen11-1} analyses
the instantaneous and average spectral efficiency of a dual-hop
network with direct link between source and destination, and a relay
node that can operate in either HD or FD mode. Interestingly, the
authors propose hybrid FD/HD relaying policies that, depending on the
channel conditions, optimally switch between the two operational modes
when the FD relay transmit power is fixed to its maximum value, as
well as when it can be reduced in order to mitigate self-interference
as needed. The FD mode only is considered in \cite{Kang14}, which aims
to maximize the signal-to-interference plus noise ratio (SINR) as the
relay transmit power varies, in the case where AF is used, the relay
has multiple transmit antennas and a single receive antenna, and
constraints on the average and maximum relay transmit power must hold.
In \cite{Alouini2016}, the maximum achievable rate and upper bounds on
the capacity are obtained when the relay node operates in DF and CF
and Gaussian inputs are considered at the source and the relay node.

The study in \cite{Zlatanov2017} is the first to derive the capacity
of the Gaussian two-hop FD relay channel with residual
self-interference, assuming the average transmit power at the source
and the relay nodes to be limited to some maximum values. The study
shows that the conditional probability distribution of the source
input, given the relay input, is Gaussian while the optimal
distribution of the relay input is either Gaussian or symmetric
discrete with finite mass points.  This result implies that a capacity
achieving scheme requires the source to know at each time instant what
the relay is transmitting. This can be realized with the aid of a
buffer at the relay, which holds the data previously transmitted by
the source and correctly decoded by the relay. The relay re-encodes
such data before forwarding it to the destination in the next
available channel use. The source can use the same encoder as the
relay, in order to predict what will be transmitted by the relay and
hence guarantee a capacity achieving transmission.

In this work, we consider a scenario similar to 
\cite{Zlatanov2017}, including a dual-hop, DF network where the relay
can operate in FD mode, and the residual self-interference is modeled
as an additive Gaussian noise, with variance proportional to the relay
transmit power.  Different from \cite{Zlatanov2017}, in this paper we
consider the case where the source does not know what symbols are
transmitted by the relay and is aware only of the transmit power
distribution adopted by the relay over a given time horizon. 
Therefore, our scenario
can accommodate the case where the relay node has to handle multiple,
simultaneous traffic flows, e.g., in-band signaling as well as data
traffic originated at the relay itself or previously received from
other sources. 
The source knowledge about the relay power is exploited in order to
optimally set the source transmit power and decide whether the relay
should operate in HD or FD. Furthermore, 
we assume a Gaussian input distribution at both source and
relay, with variance 
not exceeding a target 
maximum value. 
%The constraint on the instantaneous power is imposed by
%the limitations of the linearity of the low
%noise amplifier at the relay, and by the clipping noise of the
%analog-to-digital converter \cite{Kang14}.

Under this scenario, we formulate an optimization problem that aims at
maximizing the achievable data rate, subject to the system
constraints. We characterize different operational regions
corresponding to optimal network performance, and provide conditions
for their existence, as the values of the
system parameters vary. Our analysis led to the following major results:
\begin{itemize}

\item[{\em (i)}] The distribution of the transmit power at the relay can
  be conveniently taken as the driving factor toward the network
  performance optimization. We prove that the optimal
   distribution of such a quantity is discrete and composed of either
  one or two delta functions, depending on the target value of average
  transmit power at the source and relay. We provide the expression of
  the above distribution for the whole range of the system parameter
  values, including the channel gains and the target values for the
 average  transmit power at the source and the relay.
\item[{\em (ii)}] 
The above finding leads to the optimal communication strategy for the
network under study, which  implies either FD transmissions over an 
entire frame, or FD/HD transmissions over a fraction of the frame. 
\item[{\em (iii)}] Given the optimal transmit power distribution at
  the relay 
  and the system constraints, we  derive the optimal power
  level to be used over time at the relay and the source. Such power
  allocation policy allows the
  system to achieve the maximum data rate. We remark that our 
  policy establishes the time fractions during which the relay should
  work in FD and in HD, as well as the transmit power to be used at
  the source and the relay, given that only the average (not the instantaneous)
  relay transmit power needs to be known at the source.
\item[{\em (iv)}] We compare the results of our optimal
    power allocation to a reference FD and HD 
    scheme, where the source knows the {\em instantaneous} relay
    transmit power. Interestingly, our scheme closely approaches the
    performance of such strategy in all the considered scenarios. 
\end{itemize}

The remainder of the paper is organized as follows.  Section
\ref{sec:system} introduces our system model, while Section
\ref{sec:problem} presents the constrained optimization problem. The
optimal communication strategy and our main analytical results are
presented in Sections \ref{sec:allocation} and \ref{sec:allocation2},
for different values of the system parameters. Section
\ref{sec:results} shows the performance results, and 
Section~\ref{sec:extension} discusses how to extend the analysis to the case where
the average transmission power at the source is limited. Finally, 
Section~\ref{sec:conclusions} concludes the paper.

\section{System model\label{sec:system}}
We consider a two-hop, DF relay network including a source node $s$, a
relay $r$ and a destination $d$.  All network nodes are equipped with
a single antenna, and the relay is assumed to be FD enabled. No direct
link exists between source and destination, thus information delivery
from the source to the destination necessarily takes place through the
relay.
As far as the channel is concerned, we consider independent,
memoryless block fading channels with additive Gaussian noise, between
source and relay as well as between relay and
destination.

Source and relay operate on a {\em frame} basis, of constant duration
$T$, with $T$ being set so that channel conditions do not vary during
a frame; 
without loss of generality, in the following we set $T=1$.
In general, the following modes of operations are possible for source
and relay: (i) the source transmits while the relay receives only
(HD-RX mode); the source is inactive while the relay transmits (HD-TX mode),
(iii) the source transmits while the relay transmits and receives at
the same time (FD mode).

We remark, however, that source and relay do not need to be
synchronized on a per-symbol basis, and that the relay can handle
multiple (data or control) traffic streams originated at different
network nodes, according to any scheduling scheme of its choice. This
implies that, in order to select its operational mode, the source is
not required to be aware of the information the relay is transmitting.
We assume instead that the source has knowledge of the distribution
of the transmit power adopted by the relay across a frame.

When the relay transmits to the
destination, a residual self-interference (after analog and digital
suppression) adds up to what the relay receives from the source.
Then the signal received at the relay and destination can be written as:
\begin{eqnarray}
y_{r}&=& \sqrt{P}h_1 x_s+\nu+n_r\non
y_{d}&=& \sqrt{p}h_2 x_r+n_d \label{eq:signal}
\end{eqnarray}
where 
\begin{itemize}
\item $h_1$ and $h_2$ are the complex channel gains associated with,
  respectively, the source-relay and relay-destination links;
\item $x_s$ and $x_r$ are the input symbols transmitted by,
  respectively, the source and the relay. We assume the input at both
  source and relay to be zero-mean complex Gaussian distributed with
  unit variance. From~\eqref{eq:signal}, we have that the levels of instantaneous
  power transmitted by  source and relay, are $P|x_s|^2$ and
  $p|x_r|^2$, respectively. In the most general case, $P$ and $p$ are
  time-varying continuous random variables ranging in $[0,\Psmax]$ and
  $[0,\Prmax]$, respectively.
\item $n_r$ and $n_d$ represent zero-mean complex Gaussian noise over,
  respectively, the source-relay and the relay-destination link, with
  variance $N_0$;
\item $\nu$ represents the instantaneous residual self-interference at
  the relay. As typically done in previous studies 
  \cite{Shang2014,Shi2016,Alouini2016,Zlatanov2017}, 
we model $\nu$ as a Gaussian noise with variance 
proportional to the instantaneous transmission power at the relay, i.e.,
$|\nu|^2 = \beta p |x_r|^2$  and variance $\EE_{x_r}[|\nu|^2]=\beta p$.
In these expressions, $\beta$ denotes the self-interference attenuation factor at the
relay and $\EE_{x_r}[\cdot]$ is the expectation operator with respect to $x_r$.
Also, we remark that, as shown in \cite{Zlatanov2017},
assuming $\nu$ as a zero-mean i.i.d. Gaussian random variable
represents the worst-case linear residual self-interference model.
\end{itemize} 
  We define $f(p)$ as the probability density function of $p$, with support
in $[0, \Prmax]$. 

Finally, we consider that the average power over a frame at the source and
at the relay is constrained to given target values, denoted by $\Psav$
and $\Prav$, respectively.  The average power at the source and relay is
therefore given by:
\begin{eqnarray}
  \Prav &=& \EE_p\EE_{x_r}[p|x_r|^2] = \EE_p[p] = \int_{0}^{\Prmax} p f(p)\dd p \label{eq:f_constraint} \\
  \Psav &=& \EE_p\EE_{x_s}[P|x_s|^2] = \EE_p[P] = \int_{0}^{\Prmax} P(p) f(p)\dd p \label{eq:Ps-constraints}
\end{eqnarray}
where the expression in~\eqref{eq:Ps-constraints} is due to the fact that the source selects
  its transmission power based on its knowledge of $p$, hence $P$
  depends on $p$. 
In order to highlight this dependency,
  in the above expression and in the following, we use the $P(p)$ notation.

\section{Problem formulation\label{sec:problem}} 
In our study, we aim at determining the power allocation at
the source and relay that maximizes the achievable 
rate of the dual-hop network described above. 
To this end, we start by recalling some fundamental concepts:
\begin{itemize} 
\item[{\em (a)}] the network rate will be determined by the minimum
  between the rate achieved 
over the source-relay link and over the
relay-destination link, 
hereinafter referred to as $R_1$ and $R_2$, respectively;
\item[{\em (b)}] $R_1$ depends on the source transmit power, the
  Gaussian noise, as well as on the residual self-interference at the
  relay, which, in turn, depends on the relay transmit power;
\item[{\em (c)}] $R_2$ depends on the relay transmit power and the
  noise at the destination;
\item[{\em (d)}] the transmit power at source and relay may vary over
  time. Whenever $P>0$ and $p>0$, the relay works in FD mode, while,
  when $P>0$ and $p=0$ the relay is  receiving in HD mode. When $P=0$
  and $p>0$, the relay is transmitting in HD mode while the source is silent. 
\end{itemize}
Based on {\em (b)} and {\em (c)}, the residual self-interference
introduces a dependency 
between the performance of the first and second hop. Thus, 
in order to maximize the network rate, source and relay 
should coordinate their power allocation strategies.
In our study, we optimize the power allocation, hence the network
rate, by controlling the distribution of the transmit power, $f(p)$,  at the relay.
As a first step, we fix $f(p)$ and derive the expressions of
the rates $R_1$ and $R_2$ as detailed below.

\subsection{Optimal power distribution at the source}
Given the system model introduced above and fixed the value of $p$,
the rate on the source-relay and relay-destination links are given by
$R_1(p) = \log \left( 1+\frac{P(p) |h_1|^2}{N_0+\beta p}\right )$ and
$R_2(p) = \log\left(1+\frac{|h_2|^2}{N_0} p\right)$, respectively.
Then the average rates over a frame can be written as:
\begin{eqnarray}
  R_1 &=& \int_{0}^{\Prmax} f(p) \log \left( 1+\frac{P(p) |h_1|^2}{N_0+\beta p}\right ) \dd p\non
  R_2 &=& \int_{0}^{\Prmax} f(p)\log\left(1+vp\right)\dd p
  \label{eq:R1_z}
\end{eqnarray}
where $v \triangleq \frac{|h_2|^2}{N_0}$.

For a given distribution $f(p)$, the rate $R_1$ can be maximized with
respect to $P(p)$.  It can be shown (see Appendix~\ref{app:A}) that,
given $f(p)$, the power distribution at the source maximizing $R_1$ is
given by $P(p) = \min\left\{ \frac{\beta}{|h_1|^2} [\omega - p]^+, \Psmax\right\}$
where $\omega$ is a parameter defined as (see Appendix~\ref{app:A}): 
\[ \omega=\frac{|h_1|^2}{\beta\lambda}+\frac{N_0}{\beta}\] with
$\lambda$ being the Lagrange multiplier used in the constrained
maximization of $R_1$.
In the following, we assume that $\Psmax$ is
large enough so that
\begin{equation}
  P(p) = \frac{\beta}{|h_1|^2} [\omega - p]^+\,.
  \label{eq:Pp}
\end{equation}
We will remove this assumption and discuss the impact on the obtained
results in Section~\ref{sec:extension}.

By substituting~\eqref{eq:Pp} in~\eqref{eq:Ps-constraints}, we note
that $\omega$ has to satisfy the average transmit power constraint,
i.e.,
\begin{equation}
  \int_{0}^{\Prmax} f(p)  [\omega - p]^+\dd p =\Pcsav \triangleq
  \Psav\frac{|h_1|^2}{\beta} \,.
 \label{eq:constraint_z}
\end{equation}
Also, by substituting~\eqref{eq:Pp} in~\eqref{eq:R1_z} and by defining
$\beta_0  \triangleq\frac{\beta}{N_0} $, we get
\begin{eqnarray}
 R_1 &=&\int_{0}^{\Prmax} \hspace{-3mm} f(p)\log \left( 1\mathord{+}\beta\frac{[\omega \mathord{-} p]^+}{N_0+\beta p}\right )\dd p  \non
   & =& \int_{0}^{\Prmax} \hspace{-3mm} f(p)\log \left( 1\mathord{+}\frac{\beta_0}{1\mathord{+}\beta_0 p}[\omega \mathord{-} p]^+\right )\dd p \,.
                       \label{eq:R1_omega}
\end{eqnarray}

\subsection{Optimal power distribution at the relay}
Having expressed the source power as a function of $p$, and the rates
$R_1$ and $R_2$ as  functions of $f(p)$, we need to find the optimal
distribution $f(p)$ that maximizes the network data rate $R$. We therefore
formulate the following optimization problem, subject to the system
constraints:

%\subsection{Maximimizing the network achievable rate}
%Given the above expressions for $R_1$ and $R_2$, the goal is to maximize the system data rate over the space of distributions
%$f(z)$ meeting the constraints, i.e.,
\begin{eqnarray}
&\mbox{\bf P1:} & R=\max_{f(p)} \min\{R_1,R_2\} \quad\quad {\rm s.t.} \non
  &(a)&  R_1 = \int_{0}^{\Prmax} \hspace{-5mm}f(p) \log \left( 1\mathord{+}\frac{\beta_0 [\omega \mathord{-} p]^+}{1\mathord{+}\beta_0 p}\right)\dd p \non
  &(b)&  R_2 = \int_{0}^{\Prmax} \hspace{-3mm} f(p)\log\left(1\mathord{+}vp\right)\dd p \non
  &(c)& \int_{0}^{\Prmax} \hspace{-3mm}f(p) [ \omega \mathord{-} p]^+ \dd p = \Pcsav \non
  &(d)& \int_{0}^{\Prmax} \hspace{-3mm}p f(p)\dd p = \Prav \non
&(e)& \int_{0}^{\Prmax} \hspace{-3mm}f(p)\dd p = 1  \non
  &(f)& 0\le p \le \Prmax\nonumber
\end{eqnarray}
In the above formulation, 
\begin{itemize}
 \item constraints (a) and (b) represent the average rates achieved on the
   source-relay and relay-destination links, respectively;
 \item (c) is the average power constraint at the source;
 \item (d) is the average power constraint at the relay;
 \item (e) imposes that $f(p)$, being  a distribution, integrates to 1;
 \item (f) constraints $p$ to not exceed
   $\Prmax$.
 \end{itemize}

\section{Optimal power allocation for $\omega \geq \Prmax$\label{sec:allocation}}
In order to solve problem {\bf P1} we first consider the case $\omega \geq \Prmax$. By using such assumption in the constraints~{\em (c)}, {\em
  (d)} and~{\em (e)} of {\bf P1}, we obtain $\omega=\Pcsav+\Prav$.
Then the constraint $\omega \geq \Prmax$ implies that a solution to problem {\bf P1} exists if $\Pcsav \ge \Pc_0 = \Prmax-\Prav$.
Moreover, by using $\omega=\Pcsav+\Prav$ in~{\em (a)} and in~\eqref{eq:Pp}, we obtain
\begin{equation}
  R_1 = \log\left(1+\beta_0(\Pcsav+\Prav) \right)-\int_{0}^{\Prmax} f(p) \log(1+\beta_0 p ) \dd p
  \label{eq:R1_1}
\end{equation}
and $P(p) = \frac{\beta}{|h_1|^2}[\Pcsav+\Prav-p]$.  Since
$\log(1+cp)$, $c>0$, is a concave function of $p$ and $f(p)$ has
average $\Prav$, we can apply Lemma~\ref{lemma:1} reported in
Appendix~\ref{app:lemma1} and write:
\begin{eqnarray}
\hspace{-4mm} R_1 \hspace{-3mm}  &\le& \hspace{-3mm}  r_1^{\rm max} = \log\left(1\mathord{+}\beta_0(\Pcsav\mathord{+}\Prav)\right)\mathord{-}\frac{\Prav}{\Prmax}\log(1\mathord{+}\beta_0\Prmax) \\
 \hspace{-4mm} R_2 \hspace{-3mm}  &\ge& \hspace{-3mm}  r_2^{\rm min} = \frac{\Prav}{\Prmax}\log(1\mathord{+} v\Prmax)
\end{eqnarray}
with the equality holding when $f(p) = \left(1-\frac{\Prav}{\Prmax}\right)\delta(p) + \frac{\Prav}{\Prmax}\delta(p-\Prmax)$ where $\delta(\cdot)$ is the Dirac delta function. Similarly, by applying again Lemma~\ref{lemma:1}, we get:
\begin{eqnarray}
  R_1 &\ge& r_1^{\rm min} = \log\left(1+\beta_0(\Pcsav+\Prav)\right)-\log(1+\Prav\beta_0)  \non
  R_2 &\le& r_2^{\rm max} = \log(1+\Prav v)
\end{eqnarray}
with the equality holding when $f(p) = \delta(p-\Prav)$. Now, after
having bounded the rates $R_1$ and $R_2$, we consider the following
three cases.
\begin{enumerate}
\item If $r_2^{\rm min}\ge r_1^{\rm max}$, then $R=r_1^{\rm max}$ and
  the optimal relay power distribution is
  $f^\star(p) = \left(1-\frac{\Prav}{\Prmax}\right)\delta(p) +
  \frac{\Prav}{\Prmax}\delta(p-\Prmax)$.  Solving for $\Pcsav$ the
  inequality $r_2^{\rm min}\ge r_1^{\rm max}$, we obtain
\[\Pcsav \le \Pc_1 = \frac{1}{\beta_0}
\left[(1+\Prmax\beta_0)(1+\Prmax
  v)\right]^{\frac{\Prav}{\Prmax}}-\frac{1+\Prav\beta_0}{\beta_0}\]
 and 
\[R= \log\left(1+\beta_0(\Pcsav+\Prav)\right)-\frac{\Prav}{\Prmax}\log\left(1+\beta_0\Prmax\right)\,.\]
\item If $r_1^{\rm min}\ge r_2^{\rm max}$, then $R=r_2^{\rm max}$ and the optimal relay power distribution is 
  $f^\star(p) = \delta(p-\Prav)$.  Solving for $\Pcsav$ the inequality
  $r_1^{\rm min}\ge r_2^{\rm max}$, we get 
\begin{equation} 
\label{eq:P2}
\Pcsav \ge \Pc_2 = \Prav v\frac{1+\Prav\beta_0}{\beta_0}
\end{equation} 
and
\[R= \log\left(1+\Prav v\right)\,.\]
\item Otherwise, we find solutions for $f(p)$ such that $R=R_1=R_2$.
  Indeed, for $\Pc_1 \le \Pcsav \le \Pc_2$, problem \mbox{\bf P1} becomes:
  \begin{eqnarray}
  \mbox{\bf P2:} 
  &R&=\log\left(1+\beta_0(\Pcsav+\Prav)\right) \non
  && \hspace{-1cm}\qquad -\min_{f(p)} \int_0^{\Prmax}  f(p) \log(1+\beta_0p) \dd p  \quad \quad {\rm s.t.}\non
 % & {\rm s.t.} &\non
  &(a)& \int_0^{\Prmax}  f(p)
\log\left[(1+p \beta_0)(1+p v)\right] \dd
p=\non
&& \hspace{3cm}\log\left(1+ \beta_0(\Pcsav+\Prav)\right)\non
&(b)& \int_0^{\Prmax}  p f(p) \dd p = \Prav \non
&(c)& \int_0^{\Prmax}  f(p)\dd p = 1 \,.\nonumber
\end{eqnarray}
In this case the minimizer of the functional can be found by applying the following theorem.

\begin{theorem}
  \label{th:two_delta}
  Consider the following constrained minimization problem: 
  \begin{eqnarray}
    &\min_{f(p)}& \int_a^bf(p)\phi(p) \dd p \quad \quad {\rm s.t.}\label{eq:theorem_problem}\\
  %  &{\rm s.t.}&\non
    &(a)& \int_a^b f(p)\psi(p) \dd p = c  \non
    &(b)& \int_a^b pf(p) \dd p = m \non
    &(c)& \int_a^b f(p) \dd p = 1 \non
    &(d)& f(p) \ge 0, \forall p\in[a,b] \nonumber
  \end{eqnarray}
  where $\phi(p) = \log(1+\gamma_1 p)$, $\eta(p) = \log(1+\gamma_2
  p)$, $\psi(p) = \phi(p)+\eta(p)$, and $f(p)$ is a probability
  distribution with $p$ ranging in $[a,b]$, $a>0$. Moreover, $\gamma_1>0$,
  $\gamma_2>0$, $m\in[a,b]$ and $c$ are constant parameters.
  Then the minimizer has the following expression
     \begin{equation} \label{eq:theorem_sol}
       f^\star(p)\mathord{ =} \left\{
         \begin{array}{ll}
           \frac{p_2-m}{p_2 - a}\delta(p \mathord{-} a)+\frac{m-a}{p_2-a}\delta(p \mathord{-} p_2) & \mbox{if} \hspace{1ex}\gamma_1 \ge \gamma_2  \\
           \frac{b-m}{b-p_1}\delta(p \mathord{-} p_1)+\frac{m-p_1}{b-p_1}\delta(p \mathord{-} b) & \mbox{if} \hspace{1ex}\gamma_1  < \gamma_2 
         \end{array}\right.
     \end{equation}
  where the constants $p_1\in [a,m]$ and $p_2\in [m,b]$ are obtained by replacing \eqref{eq:theorem_sol} in the constraint {\em (a)} in \eqref{eq:theorem_problem}.
  
 \end{theorem}
 \begin{IEEEproof}
 The proof is given in Appendix~\ref{app:two_delta}.
 \end{IEEEproof}
 Through the above theorem and considering $v\ge \beta_0$, the maximizer of the rate in {\bf P2} is given by
\begin{equation}
  f^\star(p)=\frac{\Prmax-\Prav}{\Prmax-p_1}\delta(p-p_1)+\frac{\Prav-p_1}{\Prmax-p_1}\delta(p-\Prmax)
\label{eq:fp_1}
\end{equation}
where $p_1$ is obtained by replacing $f(p)$ with $f^\star(p)$ in
constraint~{\em (a)} in {\bf P2}, i.e., by solving
\begin{equation}\label{eq:p_1}
  \left[\frac{(1\mathord{+}p_1\beta_0)(1\mathord{+}p_1
      v)}{k}\right]^{\frac{\Prmax\mathord{-}\Prav}{\Prmax\mathord{-}p_1}} 
  =\frac{1\mathord{+}\beta_0(\Pcsav\mathord{+}\Prav)}{k}
\end{equation}
with $k=(1\mathord{+}\Prmax\beta_0)(1\mathord{+}\Prmax v)$.
When instead $v < \beta_0$, the maximizer of the rate in {\bf P2} is given by
\begin{equation}
  f^\star(p)=\frac{p_2-\Prav}{p_2}\delta(p)+\frac{\Prav}{p_2}\delta(p-p_2)
\label{eq:fp_2}
\end{equation}
where $p_2$ is obtained again using $f^\star(p)$ in~constraint~{\em
  (a)}, i.e., by solving
\begin{equation}\label{eq:p_2}
 \left[(1+p_2\beta_0)(1+p_2 v)\right]^{\frac{\Prav}{p_2}}=1+\beta_0(\Pcsav+\Prav)\,.
\end{equation}

\end{enumerate}
Given the optimal distribution $f^\star(p)$, which is related to the
power transmitted at the relay, the optimal power allocation at the source node can be obtained
by using~\eqref{eq:Pp}.

\begin{figure}[t]
  \centerline{\resizebox{0.5\columnwidth}{!}{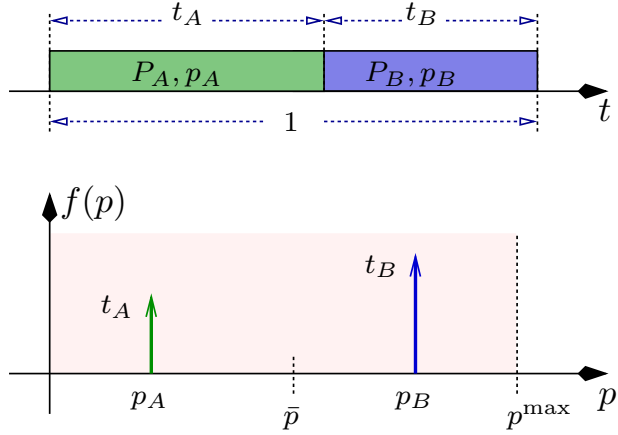}}
  \caption{Top: Optimal communication strategy during a 
    frame resulting in two phases (A and B).  
    Bottom: Optimal distribution of
    the average relay transmit power at the relay ($p$).}
  \label{fig:frame}
\end{figure}

From the above results, some important observations can be made:
\begin{itemize}
\item[{\em (i)}]   the power allocation at
the relay that leads to the maximum rate depends on the channel gain
$h_2$ through $v$ (see \eqref{eq:fp_1} and \eqref{eq:fp_2} where $p_1$
and $p_2$, given in, respectively, \eqref{eq:p_1} and
\eqref{eq:p_2}, appear). 
Similarly, the power
allocation  at the source depends on channel gain
$h_1$ (see~\eqref{eq:Pp});
\item[{\em (ii)}] 
even more importantly, the optimal power allocation
$f^\star(p)$ at the relay is {\em discrete}, 
with either one or two probability masses depending on the number of
$\delta$ functions appearing in the expression of $f^\star(p)$; 
\item[{\em (iii)}]  the above finding implies that source
and relay should operate according to a time division strategy consisting
of transmissions over either the entire frame (when $f^\star(p)$
includes one probability mass only), or two fractions of the
frame (when two  probability masses appear in $f^\star(p)$).  
Hereinafter, we will refer to such fractions as, respectively,
phase A and phase B; clearly, they reduce to one phase  when $f^\star(p)$ includes
only one probability mass. An example
where two phases exist is depicted in Figure \ref{fig:frame}(top). 
\item[{\em (iv)}]  The phases durations are given by 
the coefficients of the $\delta$ functions composing
$f^\star(p)$ (see Figure \ref{fig:frame}(bottom)). 
Note that now $p$ takes on a new meaning, as it represents the average level of transmission
power to be used at the relay during a phase of the frame.
The values of $p$, hence of the average transmission power at the relay
over each phase, are given by the arguments of the $\delta$ functions in
$f^\star(p)$. Likewise,
through~\eqref{eq:Pp}, the
average level of transmitted power at the source 
is determined by the arguments of the $\delta$  functions in $f^\star(p)$.

\end{itemize}

To summarize, Table~\ref{tab:1} reports the solution of problem
\mbox{\bf P1} for $\Pcsav \ge \Pc_0$, along with the corresponding
power allocation at the source and  relay.
\begin{table*}
  \caption{Optimal power allocation and rate for $\Pc\ge \Pc_0$ where
    $p_1$ is the solutions of~\eqref{eq:p_1} and $p_2$ is the solution
    of~\eqref{eq:p_2}.  $t_A$ and $t_B=1-t_A$ are the phases
    duration. The phases in which the relay works in HD are
    highlighted in blue\label{tab:1}}
  \begin{center}
    \begin{tabular}{|l||l|l|l||l|l|l|}
\hline
     \cellcolor{orange!20}$v\ge \beta_0$        & \multicolumn{3}{c||}{\cellcolor{orange!20}Phase A} & \multicolumn{3}{c|}{\cellcolor{orange!20}Phase B} \\ \hline
       & \multicolumn{1}{c|}{$t_A$} & \multicolumn{1}{c|}{$P_A$}   &  \multicolumn{1}{c||}{$p_A$}    & \multicolumn{1}{c|}{$t_B$} & \multicolumn{1}{c|}{$P_B$}  & \multicolumn{1}{c|}{$p_B$}  \\ \hline
    $\Pcsav \le \Pc_1$       & $\frac{\Prav}{\Prmax}$  & $\frac{\beta}{|h_1|^2}(\Pcsav+\Prav-\Prmax)$      &  $\Prmax$ &  \cellcolor{blue!20}$1-\frac{\Prav}{\Prmax}$     & \cellcolor{blue!20}$\frac{\beta}{|h_1|^2}(\Pcsav+\Prav)$  & \cellcolor{blue!20}0\\ \hline
    $\Pcsav\in(\Pc_1,\Pc_2)$     &$\frac{\Prav-p_1}{\Prmax-p_1}$ &  $\frac{\beta}{|h_1|^2}(\Pcsav+\Prav-\Prmax)$       &  $\Prmax$& $\frac{\Prmax-\Prav}{\Prmax-p_1}$& $\frac{\beta}{|h_1|^2}(\Pcsav+\Prav-p_1)$ & $p_1$ \\ \hline
    $\Pcsav \ge \Pc_2$   &  --   &  --  &  --  & 1 & $\frac{\beta}{|h_1|^2}\Pcsav$  & $\Prav$  \\ \hline \hline
     \cellcolor{orange!20}$v < \beta_0$        & \multicolumn{3}{c||}{\cellcolor{orange!20}Phase A} &
    \multicolumn{3}{c|}{\cellcolor{orange!20}Phase B} \\ \hline
                    & \multicolumn{1}{c|}{$t_A$} & \multicolumn{1}{c|}{$P_A$}   &  \multicolumn{1}{c||}{$p_A$}    & \multicolumn{1}{c|}{$t_B$} & \multicolumn{1}{c|}{$P_B$}  & \multicolumn{1}{c|}{$p_B$}  \\ \hline
    $\Pcsav \le \Pc_1$     & $\frac{\Prav}{\Prmax}$  & $\frac{\beta}{|h_1|^2}(\Pcsav+\Prav-\Prmax)$      &  $\Prmax$ &  \cellcolor{blue!20}$1-\frac{\Prav}{\Prmax}$     & \cellcolor{blue!20}$\frac{\beta}{|h_1|^2}(\Pcsav+\Prav)$  & \cellcolor{blue!20}0 \\ \hline
    $\Pcsav\in(\Pc_1,\Pc_2)$ & $\frac{\Prav}{p_2}$ & $\frac{\beta}{|h_1|^2}(\Pcsav+\Prav-p_2)$ &  $p_2$ & \cellcolor{blue!20}$1-\frac{\Prav}{p_2}$& \cellcolor{blue!20}$\frac{\beta}{|h_1|^2}(\Pcsav+\Prav)$ & \cellcolor{blue!20}$0$\\ \hline
    $\Pcsav \ge \Pc_2$ & 1 & $\frac{\beta}{|h_1|^2}\Pcsav$  & $\Prav$  &  --   &  --  &  --    \\ \hline 
  \end{tabular}
\end{center}

  \begin{center}
    \begin{tabular}{|l|l|}
    \hline
     \cellcolor{orange!20}         &\multicolumn{1}{c|}{\cellcolor{orange!20}Rate $R$} \\ \hline
    $\Pcsav \le \Pc_1$ & $\log\left(1+\beta_0(\Pcsav+\Prav)\right)-\frac{\Prav}{\Prmax}\log(1+\Prmax\beta_0)$ \\ \hline
    $\Pcsav\in(\Pc_1,\Pc_2)$; $v\ge \beta_0$ & $\log\left(1+\beta_0(\Pcsav+\Prav)\right)- \frac{\Prmax-\Prav}{\Prmax-p_1}\log\left(1+p_1\beta_0\right)- \frac{\Prav-p_1}{\Prmax-p_1}\log\left(1+\Prmax\beta_0\right)$\\ \hline
    $\Pcsav\in(\Pc_1,\Pc_2)$; $v < \beta_0$ & $\log\left(1+\beta_0(\Pcsav+\Prav)\right)- \frac{\Prav}{p_2}\log\left(1+p_2\beta_0\right)$\\ \hline
    $\Pcsav \ge \Pc_2$ & $\log(1+\Prav v)$\\ \hline
  \end{tabular}
\end{center}
\vspace{-5mm}
\end{table*}
Looking at the top tables, we remark that:
\begin{itemize}
\item for $\Pcsav \le \Pc_1$, both source and relay transmit during phase A and thus the relay operates in FD.
  In phase B, the relay is silent and only receives (HD-RX mode);
\item for $\Pc_1 \le \Pcsav \le \Pc_2$, two cases are possible. For
  $v\ge \beta_0$ the relay always operates in FD but source and relay
  use different power levels in the two phases.  Otherwise, the relay
  uses the same scheme as for $\Pcsav \le \Pc_1$, i.e., FD in phase
  A and HD-RX in phase B, but its transmit power in phase A
  should be set to $p_2$;
\item for $\Pcsav\ge \Pc_2$, the relay continuously operates in FD, 
and source and relay always transmit at their average power. 
\end{itemize}

\section{Optimal power allocation for $\omega < \Prmax$\label{sec:allocation2}}

Here, we consider  the solution of the problem {\bf P1} when
$\omega < \Prmax$. We first fix $\omega \in [0,\Prmax)$ and
then rewrite $f(p)$ as the weighted sum of two distributions, i.e.,
\begin{equation}
  f(p) = F(\omega) g(p) + [1-F(\omega)]h(p) \label{eq:f(p)split}
\end{equation}
where the distributions $g(p)$ and $h(p)$ have support in $[0,\omega]$ and
$(\omega,\Prmax)$, respectively. $F(\omega) \in [0,1]$ is the cumulative distribution function of $f(p)$, 
given by $F(\omega)=\int_0^{\omega}f(p) \dd p$.
By imposing  constraint~\eqref{eq:f_constraint}, we get
\begin{equation} \Prav = F(\omega)\int_0^\omega p g(p)\dd p + [1-F(\omega)]\int_\omega^{\Prmax} p h(p)\dd p\,.
\label{eq:Prav_g_h}
\end{equation}
If we define
\begin{equation}\label{eq:average_g}
\int_0^\omega p g(p)\dd p \triangleq \frac{G(\omega)}{F(\omega)}\,,
\end{equation}
where $0\le G(\omega)\le \Prav$, from~\eqref{eq:Prav_g_h} it immediately follows that 
\begin{equation}
  \label{eq:average_h}
  \int_\omega^{\Prmax} p h(p)\dd p
  =\frac{\Prav-G(\omega)}{1-F(\omega)} \,.
\end{equation}

For simplicity, from now on we drop the dependence on $\omega$ from
$F(\omega)$ and $G(\omega)$. Also, using the above definition, 
constraint~\eqref{eq:constraint_z} can be rewritten as
\[ \int_{0}^{\Prmax}[\omega-p]^+ f(p)\dd p = \int_0^\omega (\omega-p)F
g(p) \dd p = F\omega-G = \Pcsav \quad \quad \mbox{i.e.,}\]
\begin{equation}
  G = F\omega-\Pcsav\,.
\label{eq:G_F}
\end{equation}
We also need to impose that the averages in~\eqref{eq:average_g}
and~\eqref{eq:average_h} lie in the support of the distributions
$g(p)$ and $h(p)$, respectively. In other words,
\[ 0 \le \frac{G}{F} \le \omega; \,\,\, \omega < \frac{\Prav-G}{1-F} <
\Prmax; \,\,\, 0\le G \le \Prav \,.\]
All the above conditions can be rewritten in terms of $F$ and $\omega$
as follows: 
\begin{eqnarray}
F\le \frac{\Prav+\Pcsav}{\omega}\,; \hspace{2ex}  F< \frac{\Prmax-\Prav-\Pcsav}{\Prmax-\omega}\,;  \hspace{2ex} F\ge \frac{\Pcsav}{\omega}\,;  \hspace{2ex}\omega< \Pcsav+\Prav
\end{eqnarray}
where we recall that $\omega \in [0,\Prmax)$ and $F\in[0,1]$.

\begin{figure}[t]
 \centerline{\resizebox{0.6\columnwidth}{!}{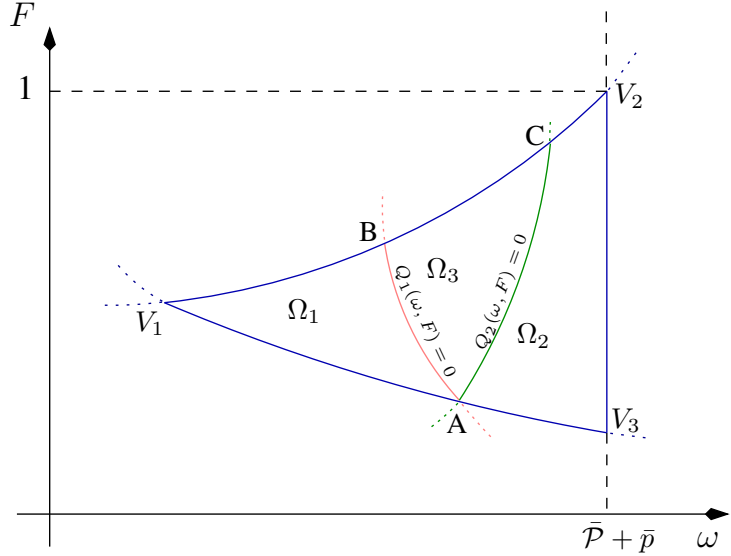}}
  \caption{A graphical representation of Region $\Omega$ and its
    subregions $\Omega_1$, $\Omega_2$ and $\Omega_3$.}
  \label{fig:triangle}
\end{figure}

Note that the condition $\omega< \Pcsav+\Prav$ implies
$\frac{\Prmax-\Prav-\Pcsav}{\Prmax-\omega}<1$ and
$\frac{\Prav+\Pcsav}{\omega} >1$.  Furthermore, in order to ensure
that  $F$ takes positive values, we must have
$\frac{\Prmax-\Prav-\Pcsav}{\Prmax-\omega}>0$, i.e.,
$\Pcsav+\Prav\le \Prmax$. Since by assumption $\omega\le \Prmax$, the above condition
is less restrictive than $\omega <\Pcsav+\Prav$. In the light of these
considerations, our conditions on $F$ reduce to
\[ \frac{\Pcsav}{\omega}\le F < \frac{\Prmax-\Prav-\Pcsav}{\Prmax-\omega} \,.\]
Clearly, a solution of the above inequalities exists if 
$\frac{\Pcsav}{\omega}\le \frac{\Prmax-\Prav-\Pcsav}{\Prmax-\omega}$, 
i.e., if $\omega \ge \frac{\Prmax\Pcsav}{\Prmax-\Prav}$.
Summarizing, these inequalities represent a region $\Omega\subset \RR^2$ defined as
\[ \Omega=\left\{(\omega,F)\in \RR^2 \Bigg| \frac{\Prmax\Pcsav}{\Prmax-\Prav} \le
    \omega \le \Pcsav+\Prav,\frac{\Pcsav}{\omega}\le F \le
    \frac{\Prmax-\Prav-\Pcsav}{\Prmax-\omega} \right\} \]
with vertices 
\begin{equation}
  \label{eq:vertices}
  V_1 = \left(\frac{\Prmax\Pcsav}{\Prmax-\Prav},1-\frac{\Prav}{\Prmax}\right);
  V_2=\left(\Pcsav+\Prav,1\right);
  V_3=\left(\Pcsav+\Prav,\frac{\Pcsav}{\Pcsav+\Prav}\right) \,.
\end{equation}
The region $\Omega$ is depicted in Figure~\ref{fig:triangle}, where the
edge $V_1$--$V_2$ has equation
$F=\frac{\Prmax-\Prav-\Pcsav}{\Prmax-\omega}$ while the edge
$V_1$--$V_3$ has equation $F=\frac{\Pcsav}{\omega}$.

It turns out that the maximization problem {\bf P1} is equivalent to
the maximization of the rate $R$ over the region $\Omega$. To this end
we substitute~\eqref{eq:f(p)split} in the expressions of the rates
$R_1$ and $R_2$ and obtain
\begin{eqnarray}
  R_2 &=& F\int_0^\omega g(p) \log(1+ p v) \dd p + (1-F)\int_\omega^{\Prmax} h(p)\log(1+ p v) \dd p \non
  &\le&
%F\int_0^\omega g(p) \log(1+p v)\dd p + (1-F)\log\left(1+v\frac{\Prav-G}{1-F}\right)  \non
%  &=&
      F\int_0^\omega g(p) \log(1+p v)\dd p + (1-F)\log\left(1+v\frac{\Prav-F\omega+\Pcsav}{1-F}\right)  \non
  &=& \widetilde{R}_2
\end{eqnarray}
where the inequality follows from Lemma~\ref{lemma:1}.
The upper bound, $\widetilde{R}_2$, is achieved for 
\begin{equation}
  \label{eq:h(p)}
  h(p) = \delta\left(p-\frac{\Pcsav+\Prav-\omega}{1-F}-\omega\right)\,.
\end{equation}
Similarly the rate $R_1$ in~\eqref{eq:R1_omega} can be rewritten as
\[R_1 = F\log(1+\beta_0\omega)-F\int_0^\omega g(p) \log(1+p\beta_0)
\dd p \,.\]
Thus, the maximization problem can be recast as
\begin{eqnarray}
  \mbox{\bf P3}  &R&=\max_{g(p), (\omega,F)\in \Omega} \min\{R_1,\widetilde{R}_2\}  \label{eq:problem_omegaF}\\
  & {\rm s.t.}&  \non
  && R_1 = F\log(1+\beta_0\omega)-F\int_0^\omega g(p) \log(1+p\beta_0) \dd p \label{eq:constraint_R1b}\\
  && \widetilde{R}_2 = F\int_0^\omega g(p) \log(1+p v)\dd p + (1-F)\log\left(1+v\frac{\Pcsav+\Prav-F\omega}{1-F}\right)\label{eq:constraint_R2b} \\
  && \int_{0}^\omega p g(p) \dd p = \omega-\frac{\Pcsav}{F}\non
  && \int_{0}^\omega g(p) \dd p = 1\nonumber
\end{eqnarray}
where the last two constraints come from~\eqref{eq:average_g},~\eqref{eq:G_F} and
the fact that $g(p)$ is a probability distribution.  In order to solve
${\bf P3}$, we first apply Lemma~\ref{lemma:1} to $R_1$ and
$\widetilde{R}_2$. By doing so, we obtain:  
\begin{eqnarray}
  R_1  &\ge&  F\log \frac{F(1+\omega \beta_0)}{F(1+\omega \beta_0)-\Pcsav\beta_0} = R_1^{\rm min} \label{eq:R1_min}\\  
             \widetilde{R}_2 &\le & F \log\left(1+v\left(\omega-\frac{\Pcsav}{F}\right)\right)
                                    +(1-F)\log\left(1+v\frac{\Pcsav+\Prav-F\omega}{1-F}\right)=\widetilde{R}_2^{\rm
                                      max}\label{eq:tilde_R2max} \,.
\end{eqnarray}
The above bounds hold with equality when
\begin{equation}
  g(p) = \delta\left(p-\omega+\frac{\Pcsav}{F}\right)\,.\label{eq:g(p)_1}
\end{equation}
  Similarly, we can write
\begin{eqnarray}
  R_1  &\le & \frac{\Pcsav}{\omega}  \log(1+\beta_0 \omega) = R_1^{\rm max} \label{eq:R1_max}\\ 
              \widetilde{R}_2 &\ge & \left(F-\frac{\Pcsav}{\omega}\right)\log(1+v\omega)
               +(1-F)\log\left(1+v\frac{\Pcsav+\Prav-F\omega}{1-F}\right)=\widetilde{R}_2^{\rm
                 min} \,.
\label{eq:tilde_R2_min} 
\end{eqnarray}
In the above expressions, equality holds for  
\begin{equation}
  g(p) = \frac{\Pcsav}{F\omega}\delta(p)+\left(1-\frac{\Pcsav}{F\omega}\right)\delta(p-\omega)\,.\label{eq:g(p)_2}
\end{equation}

\subsection{Breaking the solution space into subregions}
In order to maximize the rate over $\Omega$, we exploit the above
bounds and define the following subregions. 
\begin{itemize}
\item Let $\Omega_1 =\left\{ (\omega,F)\in\Omega \Big| R_1^{\rm min}\ge
    \widetilde{R}_2^{\rm max} \right\}$. Then in $\Omega_1$ the
  problem {\bf P3} reduces to maximizing $\widetilde{R}_2^{\rm
    max}$. The maximum rate will be denoted by $R_{\Omega_1}$.  We
  observe that $\Omega_1$ can be viewed as the set of points where
  $Q_1(\omega,F)= R_1^{\rm min}-\widetilde{R}_2^{\rm max}\ge 0$ (i.e.,
  $R_1^{\rm min}\ge \widetilde{R}_2^{\rm max}$). Then the implicit
  curve $Q_1(\omega,F)= 0$ is one of the edges of $\Omega_1$ (see
  Figure~\ref{fig:triangle}).  Also, the intersection point between
  $Q_1(\omega,F)= 0$ and the edge $V_1$--$V_3$, whose equation is
  $F=\frac{\Pcsav}{\omega}$, is
  $A=\left(\omega_A,F_A=\frac{\Pcsav}{\omega_A}\right)$. The value of
  $\omega_A$ can be computed numerically by solving
  $Q_1\left(\omega_A,F_A\right)=0$.

The intersection between $Q_1(\omega,F)= 0$ and the edge $V_1$--$V_2$,
whose equation is $F=\frac{\Prmax-\Prav-\Pcsav}{\Prmax-\omega}$, is
$B=\left(\omega_B,F_B=\frac{\Prmax-\Prav-\Pcsav}{\Prmax-\omega_B}\right)$.
The value of $\omega_B$ can be computed numerically by solving
$Q_1(\omega_B,F_B)=0$.  Moreover, we observe that the curve
$Q_1(\omega,F)$ intersects the line $\omega = \Prav+\Pcsav$ at most in
a single point. The proof is given in Appendix~\ref{app:varie}.
Finally, as shown in Appendix~\ref{app:Rminmax_omegaF},
$R_1^{\rm min}$ decreases with $\omega$ while
$\widetilde{R}_2^{\rm max}$ increases with $\omega$. Thus, we conclude
that $\Omega_1$ is located on the left of the curve $Q_1(\omega,F)= 0$
(see Figure~\ref{fig:triangle}).

\item Let
  $\Omega_2 =\left\{ (\omega,F)\in\Omega \Big| \widetilde{R}_2^{\rm
      min}\ge R_1^{\rm max} \right\}$. Then in $\Omega_2$ the problem
  {\bf P3} reduces to maximizing $R_1^{\rm max}$. The maximum rate
  achieved in this subregion will be denoted by $R_{\Omega_2}$.  We
  observe that $\Omega_2$ is given by the set of points $(\omega,F)$
  where $Q_2(\omega,F)= \widetilde{R}_2^{\rm min}-R_1^{\rm max}\ge 0$
  (i.e., $\widetilde{R}_2^{\rm min}\ge R_1^{\rm max}$). Then the
  implicit curve $Q_2(\omega,F)= 0$ is one of the edges of $\Omega_2$.
  From the results obtained in Appendix~\ref{app:Rminmax_omegaF}, we
  conclude that $Q_2(\omega,F)$ increases with $\omega$ while
  decreases with $F$. By consequence, the curve defined by the implicit
equation $Q_2(\omega,F)=0$, has positive derivative: 
\[  -\frac{\frac{\partial Q_2(\omega,F)}{\partial
      \omega}}{\frac{\partial Q_2(\omega,F)}{\partial F}}\ge 0 \,.\]
Moreover, the curve $Q_2(\omega,F)=0$ intersects the edge $V_1$--$V_3$
in $A=\left(\omega_A,F_A=\frac{\Pcsav}{\omega_A}\right)$, as it 
can be easily proven by observing that $Q_2(\omega_A,F_A)=0$.
The curve $Q_2(\omega,F)= 0$ intersects the edge $V_1$--$V_2$ in
$C=\left(\omega_C,F_C=\frac{\Pcsav+\Prmax-\Prav}{\Prmax-\omega_C}\right)$.

Note that the curve $Q_2(\omega,F)=0$ never crosses the line
$\omega=\Pcsav+\Prav$. Indeed, when $\omega=\Pcsav+\Prav$, the
expression $Q_2(\Pcsav+\Prav,F)$ does not depend on $F$ any longer.
As shown in Appendix~\ref{app:Rminmax_omegaF}, $R_1^{\rm max}$
decreases with $\omega$ while $\widetilde{R}_2^{\rm min}$ increases with
$\omega$; thus, $\Omega_2$ is located on the right of the curve
$Q_2(\omega,F)=0$ (see Figure~\ref{fig:triangle}).
  
\item Finally, let
  $\Omega_3 = \Omega \setminus (\Omega_1 \cup \Omega_2)$. The maximum
  rate achieved in $\Omega_3$ is denoted by $R_{\Omega_3}$ and can be
  obtained by maximizing the rate $R=R_1=\widetilde{R}_2$ over $g(p)$.
  To this end, we reformulate {\bf P3} as follows:
  \begin{eqnarray}
  \mbox{\bf P4} &R_{\Omega_3} &=\max_{(\omega,F)\in \Omega_3}\left[ F\log(1+\omega\beta_0)-F\min_{g(p)}\int_0^\omega g(p) \log(1+\beta_0p)\dd p\right] \label{eq:problem_region3} \\
  & {\rm s.t.}& \non
  && \int_0^\omega g(p)\left[\log(1+\beta_0p)+\log(1+vp)\right] \dd p = C(\omega,F)\non
  && \int_{0}^\omega p g(p) \dd p = \omega-\frac{\Pcsav}{F}\non
  && \int_{0}^\omega g(p) \dd p = 1\non
  && C(\omega,F) = \log(1+\beta_0\omega) +\left(1-\frac{1}{F}\right)\log\left(1+v\frac{\Pcsav+\Prav-F\omega}{1-F}\right) \label{eq:C(F)}
\end{eqnarray}
where $R_1$ is maximized with respect to $\omega, F$, and $g(p)$, and we imposed $R_1=\widetilde{R}_2$ (first constraint). 
\end{itemize}
The maximum rate over $\Omega$ is therefore given by:   
\[ R = \max \{R_{\Omega_1}, R_{\Omega_2},R_{\Omega_3}\} \,.\]
In the following, we state  the
conditions under which the three subregions exist.

\begin{figure}[t]
 \centerline{\resizebox{0.7\columnwidth}{!}{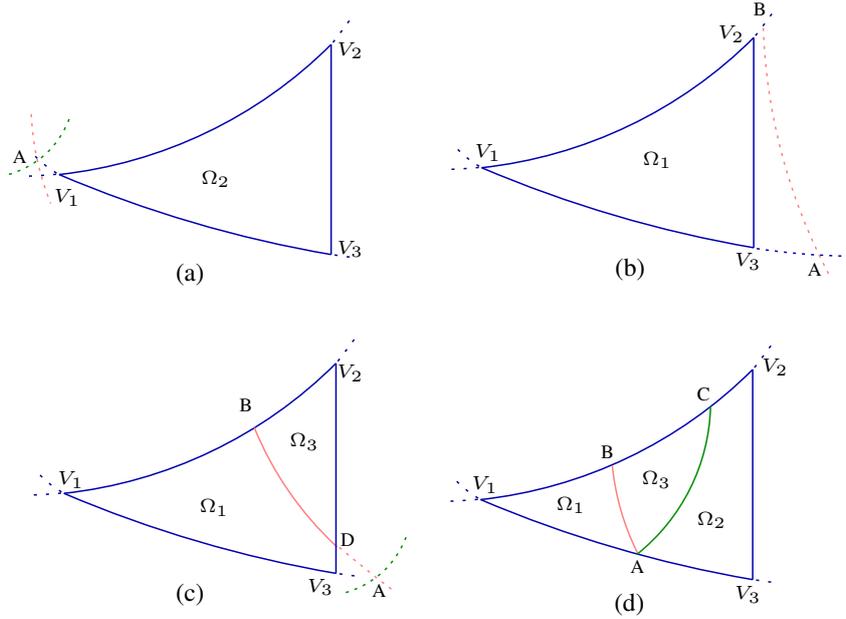}}
  \caption{A graphical representation of the subregions $\Omega_1$,
    $\Omega_2$, and $\Omega_3$ and of the cases when they exist.}
  \label{fig:triangle2}
\end{figure}

\subsection{Existence of regions $\Omega_1$, $\Omega_2$, and $\Omega_3$}
We first observe that, depending on the system parameters, the positions of the points $A$ and $B$ vary. Several cases are possible.
\begin{itemize}
\item[(a)] Point $A$ is located on the left of $V_1$, hence, outside
  $\Omega$.  Since the curve $Q_2(\omega,F)=0$ intersects the edge
  $V_1$--$V_2$ at most once, we conclude that in this case $\Omega_2=\Omega$. This
  situation is depicted in Figure~\ref{fig:triangle2}(a) and arises when $Q_2(V_1)\ge 0$.
  By solving $Q_2(V_1)\ge 0$ for $\Pcsav$, we obtain
  \[   \Pcsav \le \Pc_3 = \frac{\Prmax-\Prav}{\Prmax\beta_0}\left[(1+\Prmax v)^{\frac{\Prav}{\Prmax-\Prav}}-1\right]\,.\]
  Clearly, $\Omega_1$ and $\Omega_3$ do not exist in this case.
\item[(b)] Points $A$ and $B$ are located on the right of the points $V_3$
  and $V_2$, respectively, as depicted in Figure~\ref{fig:triangle2}(b).
  Since the curve $Q_1(\omega,F)=0$ intersects the edge $V_2$--$V_3$
  at most in a single point (as proven in Appendix~\ref{app:varie}), we
  conclude that in this case $\Omega_1=\Omega$.  The condition
  $Q_1(V_2)\ge 0$ (i.e., for which $B$ is on the right of $V_2$), solved for $\Psav$, provides
  \[   \Pcsav \ge \Prav\frac{v}{\beta_0}(1+\Prav\beta_0)=\Pc_2 \]
  while the condition $Q_1(V_3)\ge 0$ (i.e., for which $A$ is to the right of $V_3$) is equivalent to 
  \[  \Prav \log\left(1+v(\Pcsav+\Prav)\right) \le
    \Pcsav\log(1+(\Pcsav+\Prav)\beta_0)\]
  with solution  $\Pcsav \ge \Pc_{4}$. 
  Therefore, the above situation arises when
  \[\Pcsav \ge \max\{\Pc_2,\Pc_4\} \,. \]
\item[(c)] Point $A$ is located on the right of $V_3$ and $B$ is on the left of $V_2$.
  Here, only regions $\Omega_1$ and $\Omega_3$ exist, as depicted in Figure~\ref{fig:triangle2}(c). 
  This situation arises when $Q_1(V_3)\ge 0$ and $Q_1(V_2)\le 0$, i.e., for $\Pc_4 \le \Pcsav\le \Pc_2$.
  Furthermore, in this case the curve $Q_1(\omega,F)=0$ intersects the edge $V_2$--$V_3$ in $D$.  
\item[(d)] Point $A$ lies on the edge connecting $V_1$ and $V_3$. In this
  case, all regions $\Omega_1,\Omega_2$, and $\Omega_3$ exist, as depicted in Figure~\ref{fig:triangle2}(d).
  This situation happens when $\Pc_3 \le \Pcsav\le \Pc_4$.
\end{itemize}

\subsection{Maximizing the rate for varying average source power}
We consider the four cases reported in Figure~\ref{fig:triangle2}.
\begin{itemize}
\item[(a)] For $\Pcsav\le \Pc_3$ (case depicted in
  Figure~\ref{fig:triangle2}(a)),
  $R = R_{\Omega_2}=\max_{\Omega_2}R_1^{\rm max}$.  Since
  $R_1^{\rm max}$ decreases with $\omega$ and does not depend on $F$, 
  we conclude that the maximum is achieved in $V_1$. We then
  replace~\eqref{eq:g(p)_2} and~\eqref{eq:h(p)}
  in~\eqref{eq:f(p)split}, set $\omega$ and $F$ to the 
  coordinates of $V_1$, and find: 
  \[ f^\star(p) = \left(1-\frac{\Prav}{\Prmax}\right)\delta(p)+\frac{\Prav}{\Prmax}\delta(p-\Prmax)\,.\]
  Recalling that the source power is given by $P(p) = \frac{\beta}{|h_1|^2}[\omega-p]^+$, we have that
  for $p=0$ we get $P(p)=\frac{\Prmax\Psav}{\Prmax-\Prav}$, while for
  $p=\Prmax$ we have $P(p) = 0$.
  The achieved rate results to be: 
     \[ R = R_{\Omega_2}= \left(1-\frac{\Prav}{\Prmax}\right)
\log\left(1+\frac{\Prmax\Pcsav}{(\Prmax-\Prav)\beta_0}\right) \,. \]

  \item[(b)] For $\Pcsav \ge \max\{\Pc_4,\Pc_2\}$ (case depicted in Figure~\ref{fig:triangle2}(b)), $R = R_{\Omega_1}=\max_{\Omega_1}\widetilde{R}_2^{\rm max}$.
  Since $\widetilde{R}_2^{\rm max}$ increases with $\omega$ (as shown
  in Appendix~\ref{app:Rminmax_omegaF}), 
the values on the edge 
  $V_1$--$V_2$  monotonically increase  with $\omega$, and the
  values on the edge $V_1$--$V_3$
   monotonically decrease with $\omega$, we conclude that the
  maximum of $\widetilde{R}_2^{\rm max}$ is
  located on the rightmost edge of $\Omega_1$, i.e., on the edge $V_2$--$V_3$ where
  $\omega=\Pcsav+\Prav$. Once we fix $\omega$ to such value, $\widetilde{R}_2^{\rm max}|_{\omega=\Pcsav+\Prav}$
    increases with $F$. Therefore, the rate is maximized in $V_2$ and
    is given by: 
   \[ R = R_{\Omega_1} =  \log(1+v\Prav)\,. \]
   Moreover, by replacing~\eqref{eq:g(p)_1} and~\eqref{eq:h(p)} in~\eqref{eq:f(p)split} and by setting $\omega=\Pcsav+\Prav$
and $F=1$, we obtain
\[  f^\star(p) = \delta(p-\Prav)\,.\]
Since this is a single delta function, the source power can be computed for $p=\Prav$
as:  
$P(\Prav) = \frac{\beta}{|h_1|^2}[\omega-\Prav]^+ = \Psav$.

\item[(c)] For $\Pc_4 < \Pcsav <\Pc_2$, only subregions $\Omega_1$ and
  $\Omega_3$ exist, thus $R=\max\{R_{\Omega_1},R_{\Omega_3}\}$.  Let us
  first focus on $\Omega_1$. As observed before, 
  $R_{\Omega_1}=\max_{\Omega_1}\widetilde{R}_2^{\rm max}$ where 
  $\widetilde{R}_2^{\rm max}$ increases with $\omega$, thus $\widetilde{R}_2^{\rm max}$  is
  maximized on the edge $B$--$D$ and on the segment $D$--$V_3$ (where
  $\omega=\Pcsav+\Prav$). However, as mentioned for
  $\omega=\Pcsav+\Prav$, $\widetilde{R}_2^{\rm max}$ increases with
  $F$. It follows that the maximum must lie on the edge $B$--$D$.

  As for the subregion $\Omega_3$, the maximum achievable rate is given
  by the solution of {\bf P4}, which can be solved by using
  Theorem~\ref{th:two_delta}. We have that:
  \begin{itemize}
  \item if $v\ge \beta_0$, as shown in Appendix~\ref{app:region3}, 
    $R_{\Omega_3}$ lies on the edge $B$--$D$.
    Thus, $R=R_{\Omega_1}=R_{\Omega_3}$ and $R$ can be computed by solving
    $R=\max_{Q_1(\omega,F)=0}R_1^{\rm min}$, which is convex, hence,
    easy to be solved. Let
    $(\omega^\star,F^\star)$ be the point where the rate is
    maximized, then the corresponding function $f^\star(p)$ is given by combining~\eqref{eq:h(p)}
    with $g(p) = \delta(p-\omega+\Pcsav/F)$, i.e., 
    \begin{equation}
      \label{eq:f_star}
      f^\star(p) =
      F^\star\delta(p-\omega^\star+\Pcsav/F^\star)+(1-F^\star)\delta\left(p-\frac{\Pcsav+\Prav-\omega^\star}{1-F^\star}-\omega^\star\right) \,;
    \end{equation}
  \item otherwise, as shown in Appendix~\ref{app:region3}, the problem
    can be solved numerically and the rate is maximized in
    $V_2=(\Pcsav+\Prav,1)$. The optimal distribution of the transmission power at the
    relay is then given by: 
    \begin{equation}
      \label{eq:f_star_v2}
      f^\star(p) =
      \frac{p_2-\Prav}{p_2}\delta(p)+\frac{\Prav}{p_2}\delta (p-p_2) \,.
    \end{equation}
  \end{itemize}
  
\item[(d)] When $\Pc_3 < \Pcsav < \Pc_4$, the situation is depicted in
  Figure \ref{fig:triangle2}(d) where all three subregions exist. In
  subregion $\Omega_1$, following the same rational as in case (c), we
  conclude that the rate $R_{\Omega_1}$ lies on the edge $B$--$A$. In
  subregion $\Omega_2$, the rate is
  $R_{\Omega_2}=\max_{\Omega_2}R_1^{\rm max}$. Since $R_1^{\rm max}$
  does not depend on $F$, it decreases with $\omega$, and the implicit
  curve $Q_2(\omega,F)=0$ is monotonically increasing, we conclude
  that $R_{\Omega_2}$ is obtained when operating in $A$. Hence, $R_{\Omega_2}\le R_{\Omega_1}$.
  With regard to  subregion $\Omega_3$, the maximum achievable rate is given
  by the solution of {\bf P4}, which can be solved by using
  Theorem~\ref{th:two_delta}. We have that:
  \begin{itemize}
    \item if $v\ge \beta_0$, as observed for case (c), $R_{\Omega_3}$ lies on the edge $B$--$A$.
    Thus, $R=R_{\Omega_1}=R_{\Omega_3}$ and $R$ can be computed by solving
    $R=\max_{Q_1(\omega,F)=0}R_1^{\rm min}$;  
     \item else, similarly to the previous case (see
       Appendix~\ref{app:region3}), 
the problem can be solved numerically
 and the optimum is located in $A=(\omega_A,F_A)$. 
The optimal distribution of the transmission power at the relay is
given by: 
    \begin{equation}
      \label{eq:f_star_v3}
      f^\star(p) =
      F_A\delta(p)+(1-F_A)\delta\left(p-\frac{\Prav}{1-F_A}\right) \,.
    \end{equation}
   \end{itemize}

\begin{table}
  \caption{Optimal power allocation and rate for $\Pc < \Pc_0$ where $p_2$ is the solutions of~\eqref{eq:constraint_region3_2} for $F=1$ and $\omega =\Pcsav+\Prav$.
    $t_A$ and $t_B=1-t_A$ denote the phases duration. The phases in which the relay works in HD are highlighted in blue  \label{tab:2}}
  \begin{center}
    {\small
      \begin{tabular}{|c|l||l|l|l||l|l|l|}
    \hline
      \multicolumn{2}{|c|| }{\cellcolor{orange!20}$v \ge \beta_0$}    & \multicolumn{3}{c||}{\cellcolor{orange!20}Phase A} &  \multicolumn{3}{c|}{\cellcolor{orange!20}Phase B} \\ \hline
         \multicolumn{2}{|c|| }{}         & \multicolumn{1}{c|}{$t_A$} & \multicolumn{1}{c|}{$P_A$}    & \multicolumn{1}{c||}{$p_A$}    & \multicolumn{1}{c|}{$t_B$} & \multicolumn{1}{c|}{$P_B$}  & \multicolumn{1}{c|}{$p_B$}  \\ \hline
    (a) &$\Pcsav \le \Pc_3$                  &  \cellcolor{blue!20}$\frac{\Prav}{\Prmax}$   &    \cellcolor{blue!20}$0$   &  \cellcolor{blue!20} $\Prmax$  & \cellcolor{blue!20}$1-\frac{\Prav}{\Prmax}$ &  \cellcolor{blue!20}$\frac{\beta}{|h_1|^2}\frac{\Prmax\Pcsav}{\Prmax-\Prav}$  &    \cellcolor{blue!20}0 \\ \hline
    (b) & $\Pcsav \ge \max\{\Pc_2,\Pc_4\}$    & --   & --   & --  & 1 & $\frac{\beta}{|h_1|^2}\Pcsav$ & $\Prav$   \\ \hline
    (c), (d) &$\Pcsav\in(\Pc_3,\max\{\Pc_2,\Pc_4\})$   & \cellcolor{blue!20}$1-F^\star$ & \cellcolor{blue!20}0  & \cellcolor{blue!20}$\frac{\Pcsav+\Prav-F^\star\omega^\star}{1-F^\star}$ & $F^\star$ &  $\frac{\beta}{|h_1|^2}\frac{\Pcsav}{F^\star}$  &  $\omega^\star-\frac{\Pcsav}{F^\star}$ \\ \hline
    \hline
      \multicolumn{2}{|c|| }{\cellcolor{orange!20}$v < \beta_0$}    & \multicolumn{3}{c||}{\cellcolor{orange!20}Phase A} &  \multicolumn{3}{c|}{\cellcolor{orange!20}Phase B} \\ \hline
         \multicolumn{2}{|c|| }{}         & \multicolumn{1}{c|}{$t_A$} & \multicolumn{1}{c|}{$P_A$}    & \multicolumn{1}{c||}{$p_A$}    & \multicolumn{1}{c|}{$t_B$} & \multicolumn{1}{c|}{$P_B$}  & \multicolumn{1}{c|}{$p_B$}  \\ \hline
    (a) &$\Pcsav \le \Pc_3$                  &  \cellcolor{blue!20}$\frac{\Prav}{\Prmax}$   &    \cellcolor{blue!20}$0$   &  \cellcolor{blue!20} $\Prmax$  & \cellcolor{blue!20}$1-\frac{\Prav}{\Prmax}$ &  \cellcolor{blue!20}$\frac{\beta}{|h_1|^2}\frac{\Prmax\Pcsav}{\Prmax-\Prav}$  &    \cellcolor{blue!20}0 \\ \hline
    (b) & $\Pcsav \ge \max\{\Pc_2,\Pc_4\}$  & 1 & $\frac{\beta}{|h_1|^2}\Pcsav$ & $\Prav$   & --   & --   & --    \\ \hline
    (c) &$\Pcsav\in(\Pc_4,\max\{\Pc_2,\Pc_4\})$  & $\frac{\Prav}{p_2}$ & $\frac{\beta(\Pcsav+\Prav-p_2)}{|h_1|^2}$  &  $p_2$ & \cellcolor{blue!20}$1-\frac{\Prav}{p_2}$ & \cellcolor{blue!20}$\frac{\beta(\Pcsav+\Prav)}{|h_1|^2}$  &  \cellcolor{blue!20}0    \\ \hline
    (d) &$\Pcsav\in(\Pc_3,\Pc_4]$   & \cellcolor{blue!20}$1-F_A$  & \cellcolor{blue!20} $0$ & \cellcolor{blue!20}$\frac{\Prav}{1-F_A}$&\cellcolor{blue!20} $F_A$ & \cellcolor{blue!20}$\frac{\beta\Pcsav}{|h_1|^2F_A}$ & \cellcolor{blue!20}$0$   \\ \hline    
      \end{tabular}}
%\end{center}
    \smallskip
    
%  \begin{center}
{\small
    \begin{tabular}{|c|l|l|}
    \hline
    \multicolumn{2}{|c| }{\cellcolor{orange!20}}    & \multicolumn{1}{c|}{\cellcolor{orange!20}Rate $R$} \\ \hline
    (a)  & $\Pcsav \le \Pc_3$ & $(1-\frac{\Prav}{\Prmax})\log\left(1+\frac{\Prmax\Pcsav \beta_0}{\Prmax-\Prav}\right)$ \\ \hline
    (b)  & $\Pcsav \ge \max\{\Pc_2,\Pc_4\}$ & $\log(1+\Prav v)$ \\ \hline
    (c), (d) & $\Pcsav\in(\Pc_3,\max\{\Pc_2,\Pc_4\})$; $v \ge \beta_0$ & $F^\star\log\left(\frac{F^\star(1+\omega^\star\beta_0)}{F^\star(1+\omega^\star\beta_0)-\Pcsav\beta_0}\right)$\\ \hline
    (c) & $\Pcsav\in(\Pc_4,\max\{\Pc_2,\Pc_4\})$, $ v < \beta_0 $ & $\log(1+\beta_0(\Pcsav+\Prav))-\frac{\Prav}{p_2}\log(1+\beta_0p_2)$\\ \hline
    (d) &$\Pcsav\in(\Pc_3,\Pc_4]$; $v <\beta_0$  & $F_A\log(1+\beta_0\omega_A)$ \\ \hline
  \end{tabular}}
  \end{center}
\end{table}

As done before, we use the obtained power probability density function
of the transmit power at the relay, to derive
the optimal power allocation at the source node using~\eqref{eq:Pp}.
In Table~\ref{tab:2}, we report our results highlighting the power
allocation at both source and relay, the phases duration, and the data
rate for the different cases analyzed above. By looking at the top tables, we
can make the following observations: 
\begin{itemize}
\item for $\Pcsav \le \Pc_3$, the source transmits during phase B only 
  (i.e., the relay operates in HD-RX mode) while in phase A the relay
  operates in HD  transmitting at its maximum power (HD-TX mode);
\item for $\Pcsav \ge \max\{\Pc_2,\Pc_4\}$, the relay operates in FD
  mode for the whole frame and both source and relay transmit at their
  average power;
\item for $\Pcsav\in(\Pc_3,\max\{\Pc_2,\Pc_4\})$ and $v \ge \beta_0$, 
  the relay works in  HD-TX in phase A and in FD mode during phase B;
\item for $\Psav\in(\Pc_4,\max\{\Pc_2,\Pc_4\})$ and $ v < \beta_0 $, the
  relay works in  FD in phase A and in HD-RX mode during phase B;
\item for $\Pcsav\in(\Pc_3,\Pc_4]$ and $v <\beta_0$, the
  relay works in HD-TX mode during phase A and in HD-RX in phase B; 
thus, this case corresponds to the traditional HD mode.
\end{itemize}

\end{itemize}

\section{Results\label{sec:results}}
\insertfig{0.8}{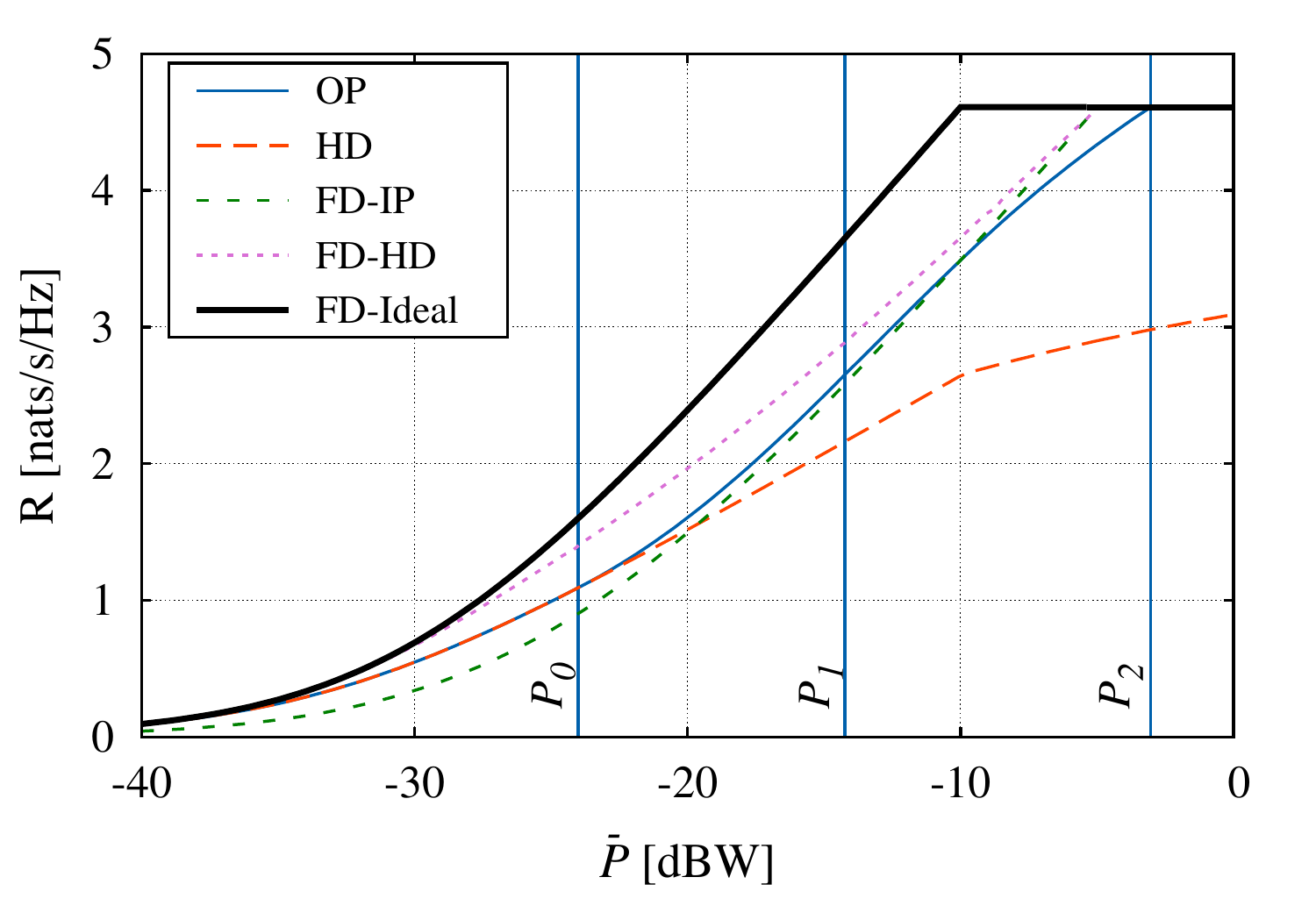}{Achieved rate vs. $\Psav$, for
  $\Prav=-10$\,dBW, $\Prmax=-7$\,dBW and $\beta=-135$\,dB.}{fig:figure135}

\insertfig{0.8}{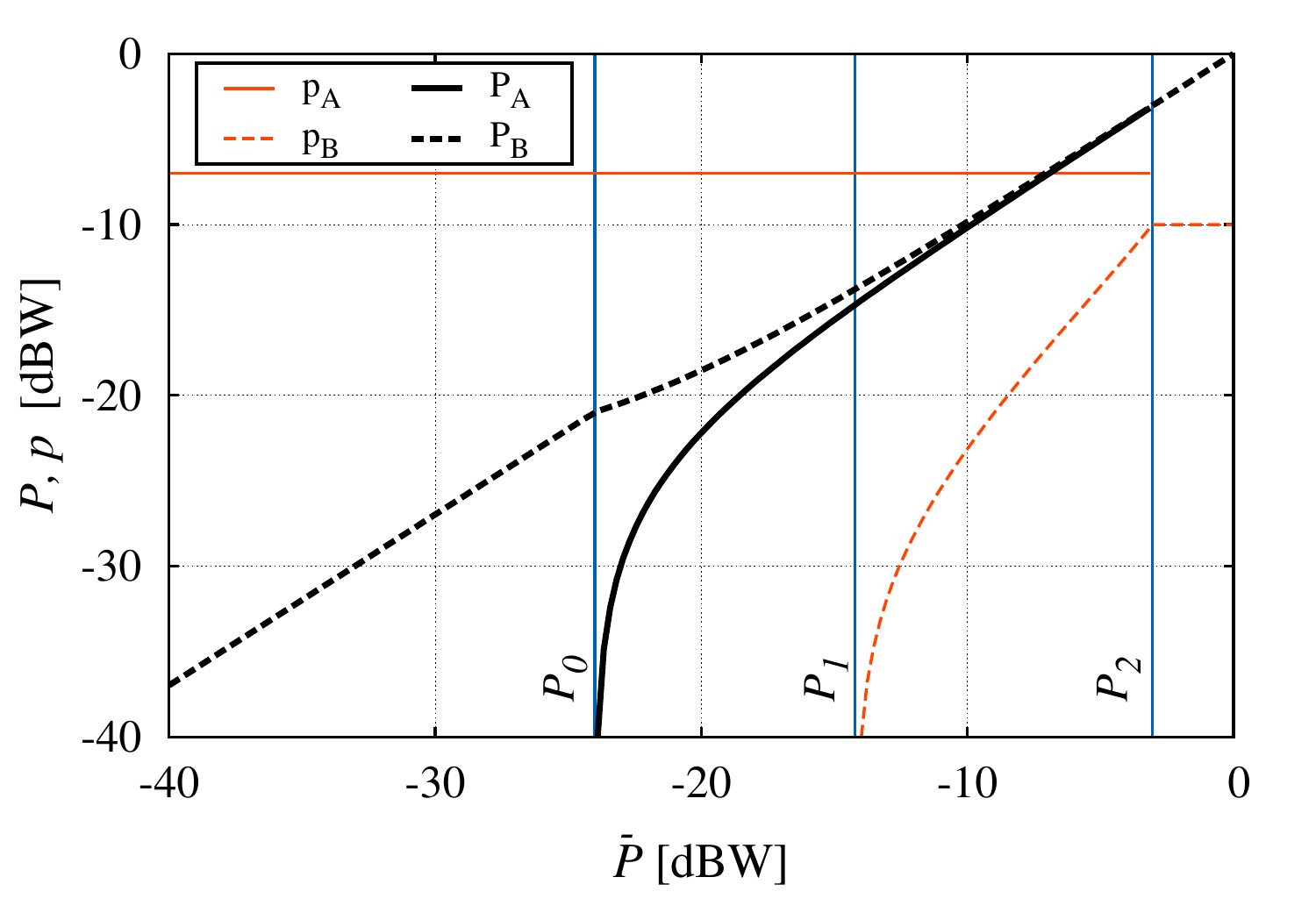}{Optimal source and relay
  transmit power for phase A (solid lines) and phase B (dashed lines), for
  the same scenario as in Figure~\ref{fig:figure135}.}{fig:figure135_slot_a}

\insertfig{0.8}{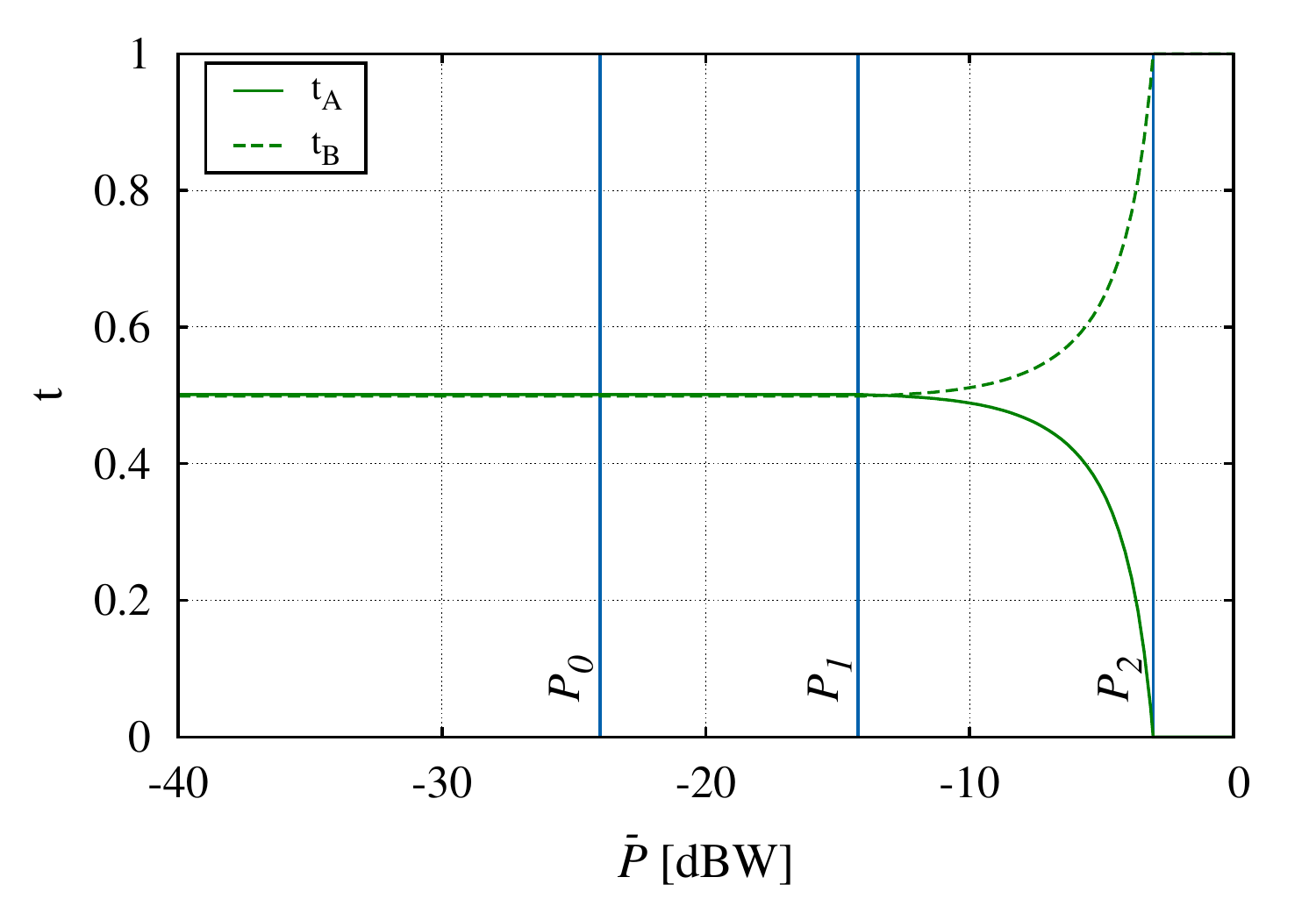}{Phase durations $t_A$ (solid
  line) and $t_B$ (dashed line), 
for the same scenario as in   Figure~\ref{fig:figure135}.}{fig:figure135_slot_b}

\insertfig{0.8}{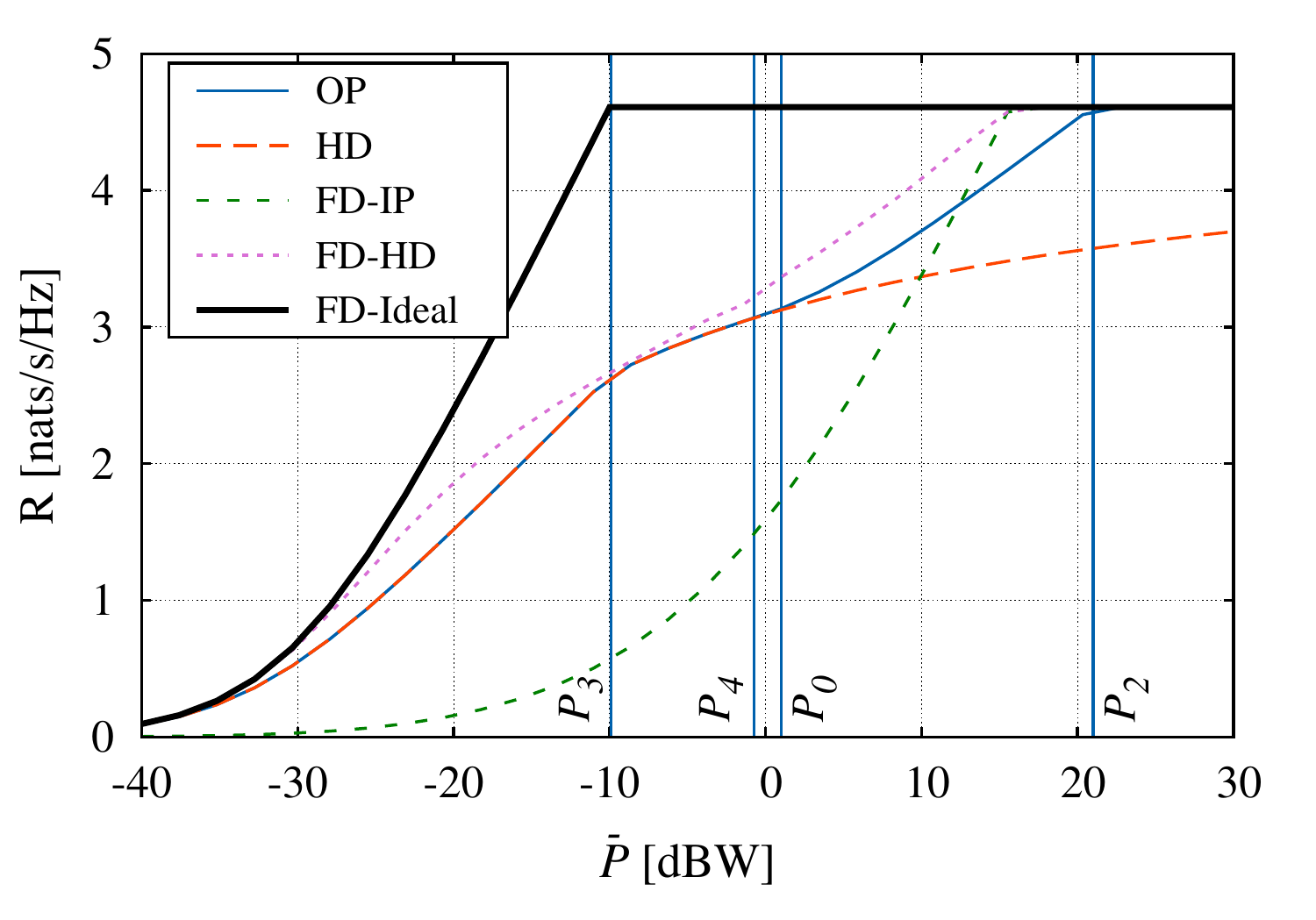}{Achieved rate vs. $\Psav$, for
  $\Prav=-10$\,dBW, $\Prmax=-7$\,dBW and $\beta=-110$\,dB.}{fig:figure110}

\insertfig{0.8}{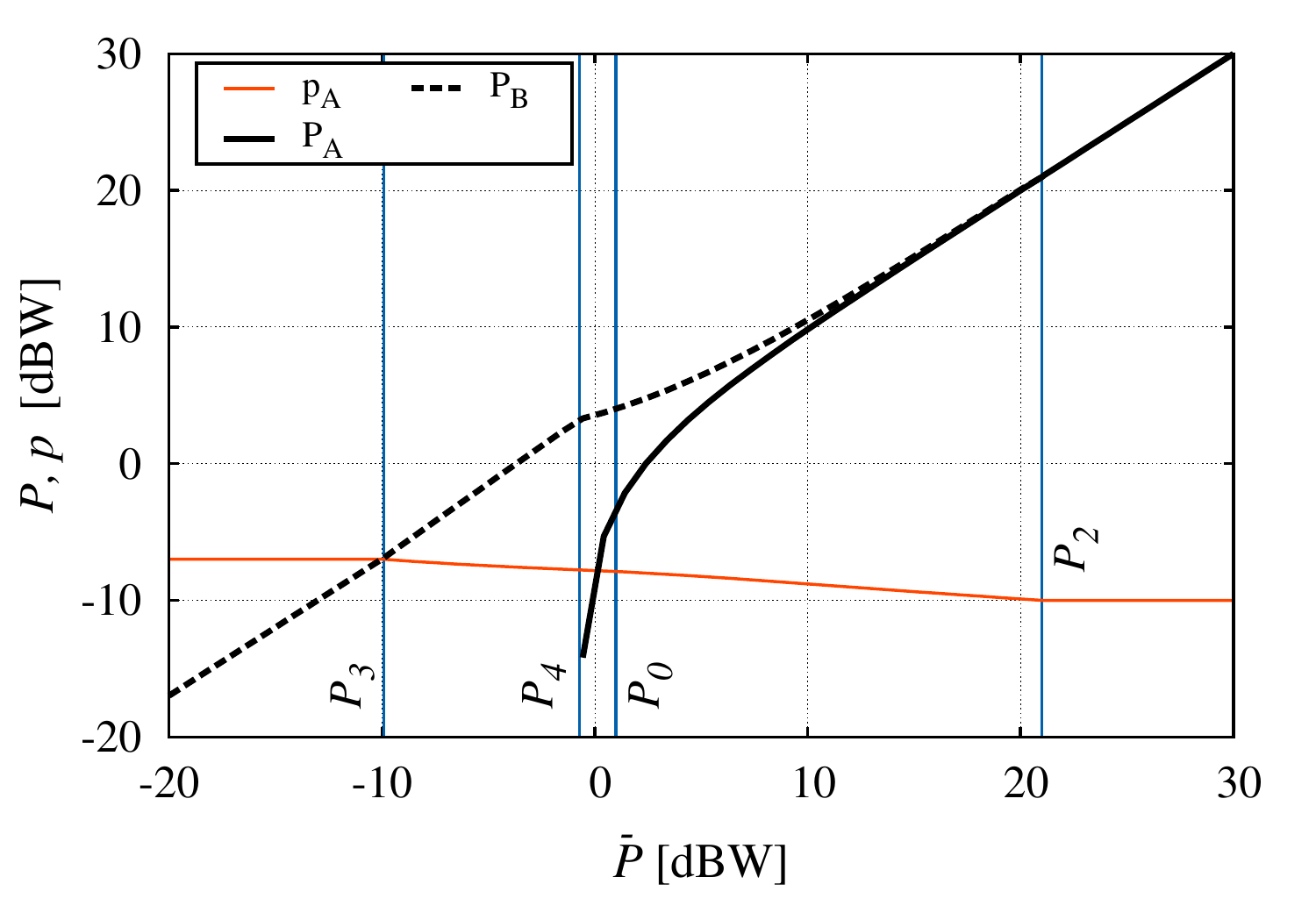}{Optimal source and relay
  transmit power for phase A (solid lines) and phase B (dashed lines), for
  the same scenario as in Figure~\ref{fig:figure110}.}{fig:figure110_slot_a}

\insertfig{0.8}{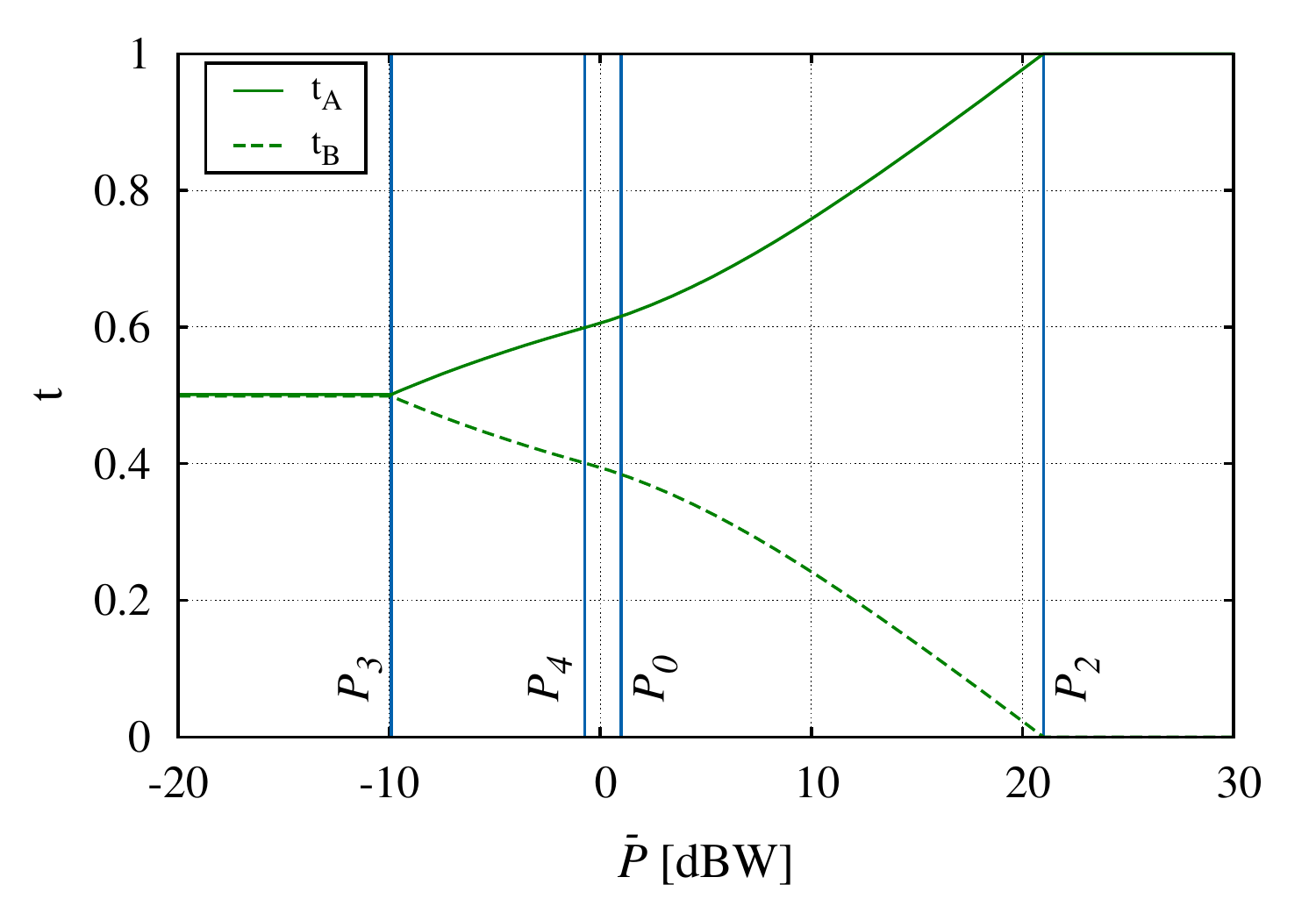}{Phase durations $t_A$ (solid
  line) and $t_B$ (dashed line), for the same scenario as in
  Figure~\ref{fig:figure110}.}{fig:figure110_slot_b}

\insertfig{0.8}{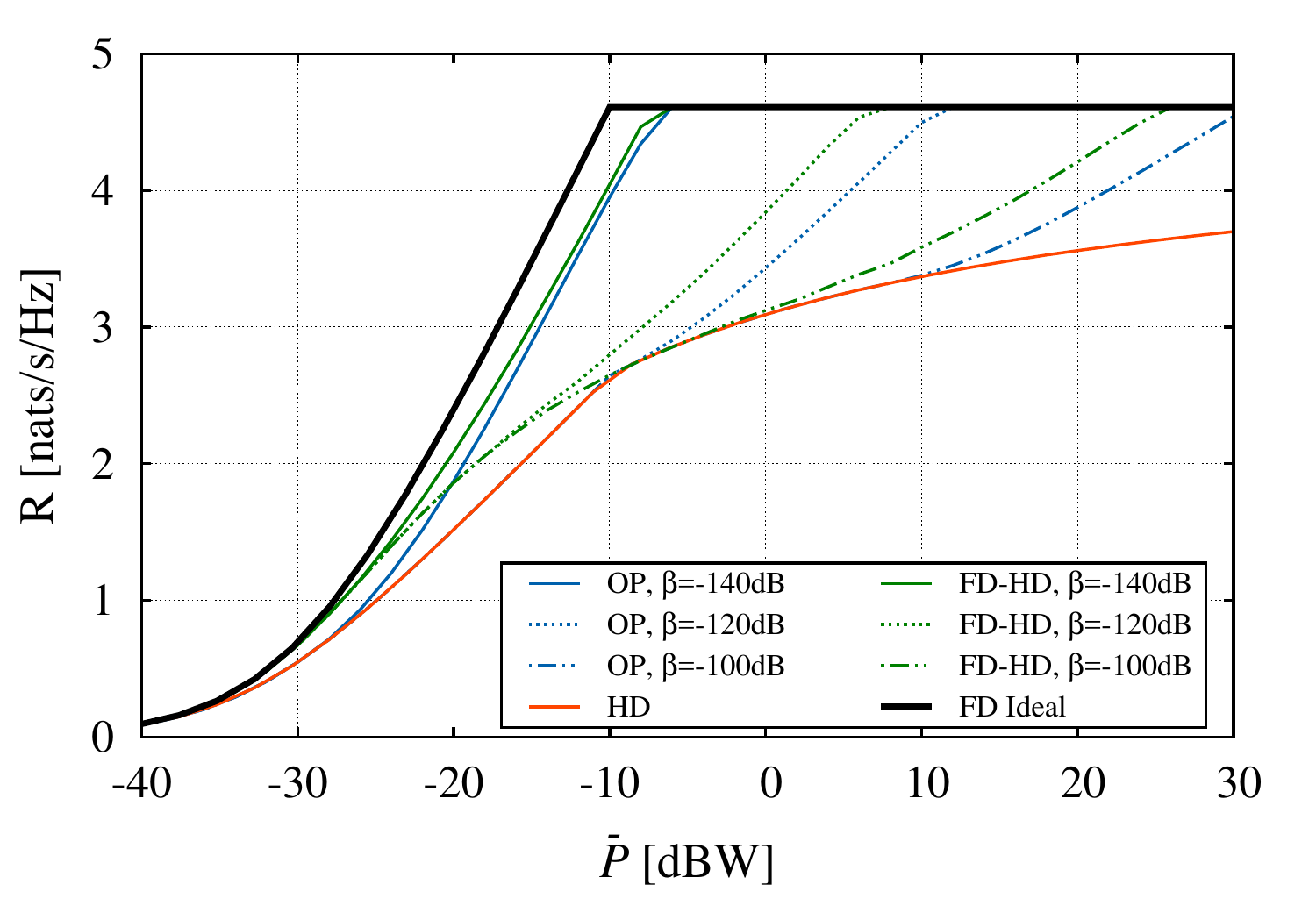}{Achieved rate vs. $\Psav$, for
  $\Prav=-10$\,dBW, $\Prmax=-7$\,dBW, and different values of
  $\beta$.}{fig:figure3}

\insertfig{0.8}{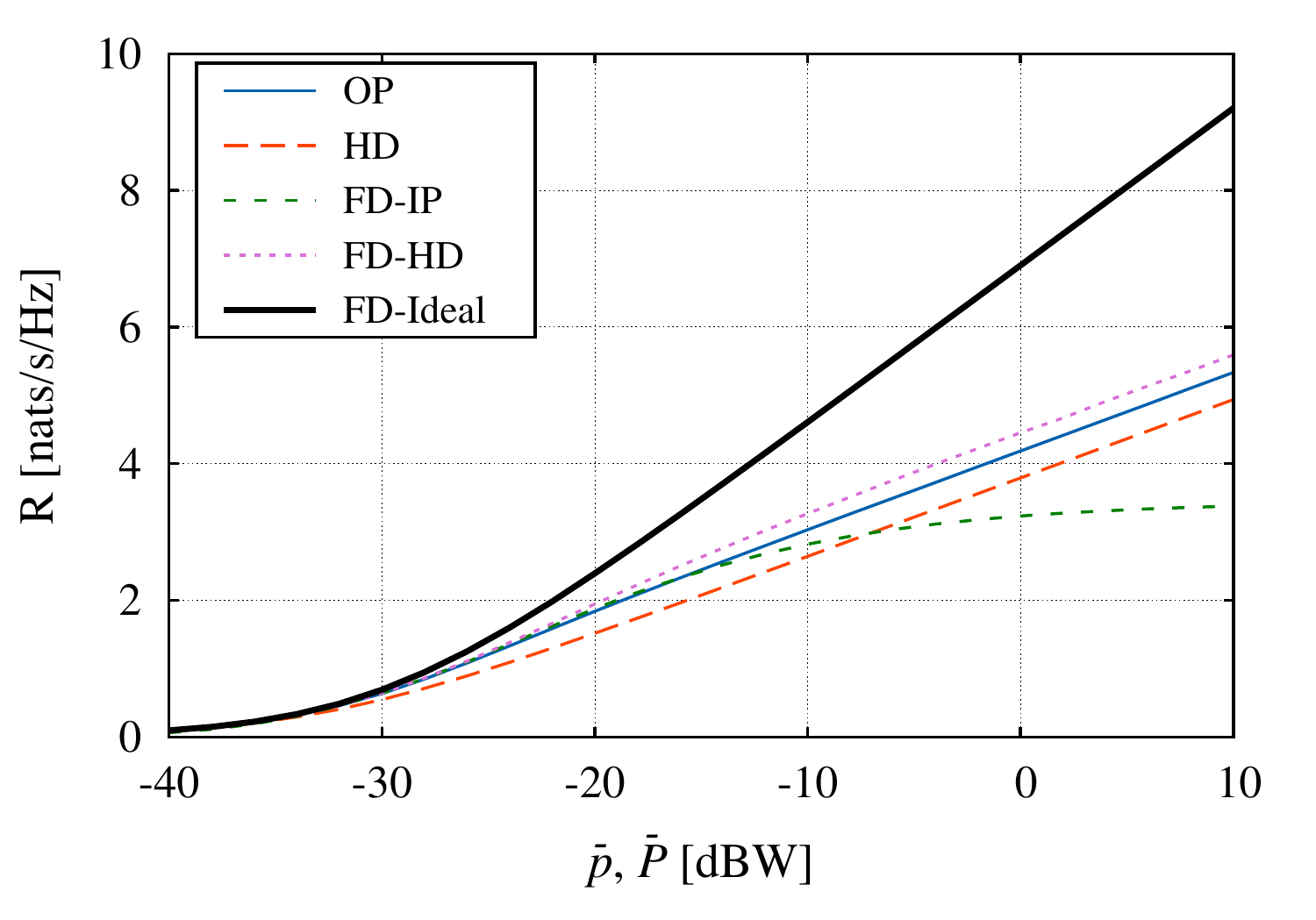}{Achieved rate vs. $\Psav=\Prav$, for
  $\Prmax=\Prav+3$\,dB and $\beta=-130$\,dB.}{fig:figure_pP_3}

\insertfig{0.8}{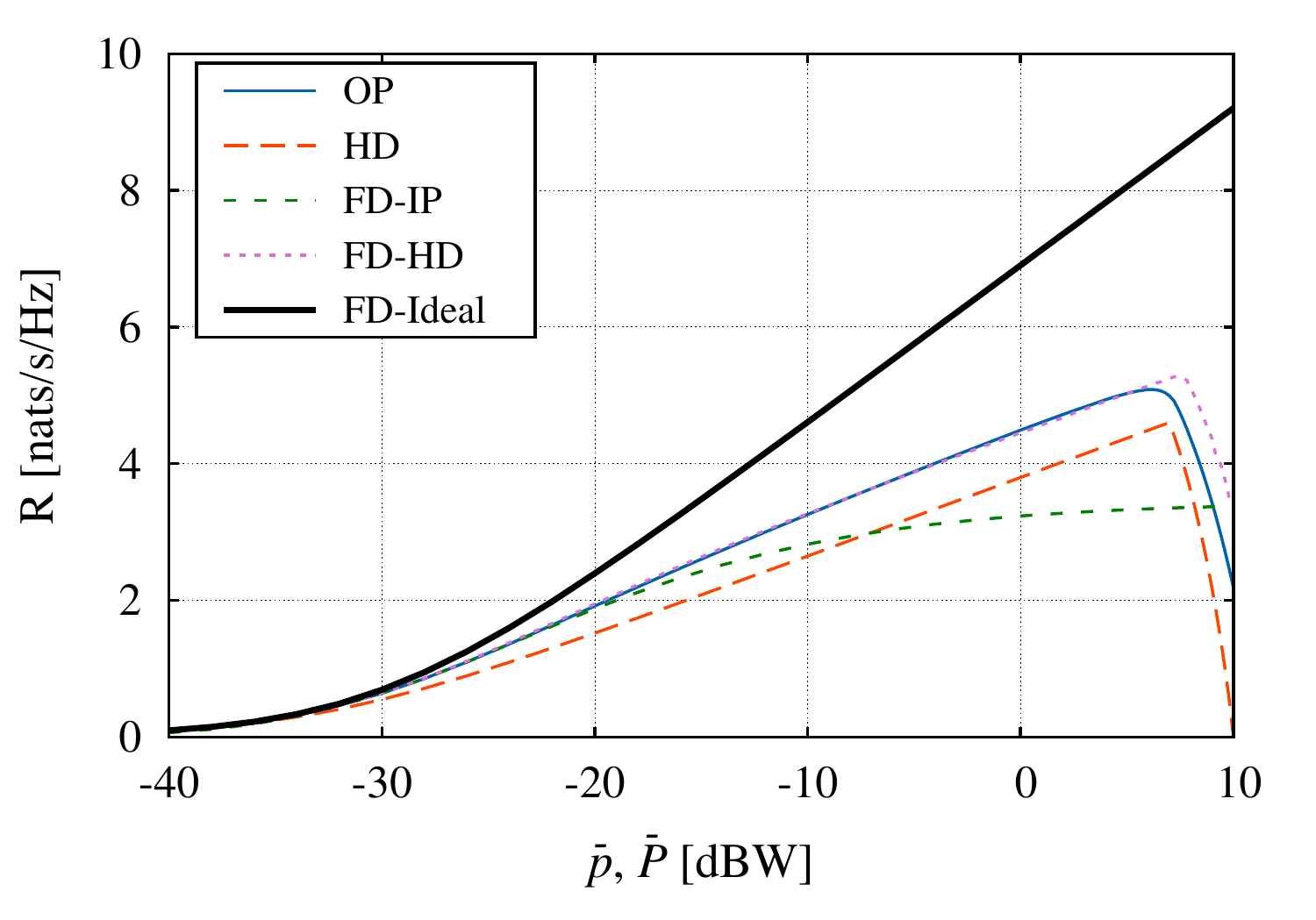}{Achieved rate vs. $\Psav=\Prav$, for
  $\Prmax=10$\,dBW and $\beta=-130$\,dB.}{fig:figure_pP_40}

We compare the performance of our proposed scheme against the ideal
full duplex communication scheme (in the following referred to as ``FD
Ideal'') where the relay does not suffer from self-interference. The
expression of the corresponding rate is
\[ R_{\rm FD-Ideal}=\min\left\{\log\left(1+\frac{\Psav
        |h_1|^2}{N_0}\right),\log\left(1+ \frac{\Prav
        |h_2|^2}{N_0}\right)\right\}\] which is also reported
in~\cite[eq.(38)]{Zlatanov2017}.  We then consider the full duplex
scheme (referred to as ``FD-IP'') where the source is aware of the
instantaneous power (IP) at which the relay transmits. In FD-IP, the
source always transmits with average power $\Psav$ while the relay
transmits with average power $\Prav$. We stress that, unlike FD-IP, our
scheme only requires the knowledge at the source of the average power
used at the relay. The expression of the rate achieved by FD-IP is:
\begin{eqnarray}  R_{\rm FD-IP}&=&\min\left\{\int_{-\infty}^{+\infty}\log\left(1+\frac{\Psav |h_1|^2}{N_0+\beta x^2}\right) \frac{\ee^{-x^2/(2\Prav)}}{\sqrt{2\pi \Prav}}\dd x, \log\left(1+\frac{\Prav |h_2|^2}{N_0} \right)\right\}\,.
\end{eqnarray}
Furthermore, we compare our solution to the conventional half duplex scheme (named
``HD''), for which the rate is given by  
\begin{eqnarray}
R_{\rm HD}&=&\max_{\Prav/\Prmax \le t \le 1}\min \left\{(1\mathord{-}t)\log\left(1\mathord{+}\frac{|h_1|^2\Psav}{(1\mathord{-}t)N_0}\right), t\log\left(1+\frac{\Prav |h_2|^2}{t N_0}\right)\right\}
\end{eqnarray}
where the relay always operates in half duplex and its transmit power
is limited to $\Prmax$.  This scheme implies that the communication is
organized in two phases of duration $t$ and $1-t$, respectively.

Finally, we consider a hybrid communication scheme named FD-HD
where the relay leverages on FD-IP or on HD, depending on which
operational mode provides the highest rate. Specifically, FD-HD is organized
in the following three phases: (A)
the source transmits at power $P_A$ for a time fraction $t_A$ while
the relay is silent; (B) the source is silent and the relay transmits
at power $p_B$ for a time fraction $t_B$; (C) the relay operates in
FD, source and relay transmit at power $P_C$ and $p_C$, respectively, for a
time fraction $t_C$. As in FD-IP, the source has knowledge of the
instantaneous power used by the relay.
The achieved rate is given by: 
\begin{eqnarray}  R_{\rm FD-HD}
  &=&\max_{\substack{t_A,t_B,t_C \\ P_A,P_C \\ p_B,p_C}} \min\left\{t_A\log\left(1+\frac{P_A|h_1|^2}{N_0}\right) +t_C\int_{-\infty}^{+\infty}\log\left(1+\frac{P_C |h_1|^2}{N_0+\beta x^2}\right) \frac{\ee^{-x^2/(2p_C)}}{\sqrt{2\pi p_C}}\dd x, \right. \non
  && \left. t_B\log\left(1+\frac{p_B|h_2|^2}{N_0}\right)+ t_C\log\left(1+\frac{p_C |h_2|^2}{N_0} \right)\right\}
\end{eqnarray}
where the first argument of the $\min$ operator represents the rate
achieved on the source-relay link, the second one represents the
rate achieved on the relay-destination link, and the following
constraints must hold: $t_A+t_B+t_C=1$, $t_AP_A+ t_CP_C=\Psav$,
$t_Bp_B+ t_Cp_C=\Prav$, and $p_B,p_C \le \Prmax$.

In order to evaluate the performance of our solution against the above
schemes, we consider a scenario similar to that employed
in~\cite{Zlatanov2017} where the source-relay and relay-destination
distances are both set to $d=500$\,m, the signal carrier frequency is
$f_c=2.4$\,GHz and the path loss is given by
$|h_1|^2=|h_2|^2 = \left(\frac{c}{4\pi f_c}\right)^2 d^{-\alpha}$,
with $\alpha=3$. Considering an additive noise with power spectral
density -204\,dBW/Hz and a bandwidth $B=200$\,kHz,  the noise power
at both relay and destination receivers is about $N_0= -151$\,dBW. 
Note that, for this setting, we have $v =|h_1|^2/N_0 \approx 30$\,dB.
  
Figure~\ref{fig:figure135} compares the rate of our optimal power
allocation scheme, labeled ``OP'', against the performance of FD-Ideal, FD-IP, FD-HD and HD, for $\Prav=-10$\,dBW, $\Prmax=-7$\,dBW and
$\beta=-135$\,dB. Since $\beta_0 = \beta/N_0 \approx 16$\,dB, the
results we derived for $v>\beta_0$ apply. Let $P_i =\Pc_i\frac{\beta}{|h_1|^2}$,
for $i=0,\ldots,4$.  For the parameters used in this example, the value
of the thresholds $P_i$ ($i=0,\ldots, 4$) are: $P_0=-24$\,dBW, $P_1=-14.23$\,dBW,
$P_2=-3.04$\,dBW, $P_3=-9.92$\,dBW, and $P_4=-20.56$\,dBW.  The thresholds $P_3$
and $P_4$ are meaningful only if lower than $P_0$ (see
Section~\ref{sec:allocation2}), thus they are not shown in the
figure. The achieved rates are depicted as functions of the average
transmit power at the source, $\Psav$.  For $\Psav \ge P_0$, the
results obtained in Section~\ref{sec:allocation} hold. Accordingly,
the plot highlights three operational regions corresponding to
$P_0 \le\Psav\le P_1$, $P_1<\Psav\le P_2$, and $\Psav > P_2$,
respectively.  Instead, for $\Psav < P_0$ (see
Section~\ref{sec:allocation2}), we have a single operational
region only, since $P_3>P_0$ and $P_4>P_0$.  We observe that all
communication strategies are outperformed by FD-Ideal, which assumes no
self-interference at the relay. Also, FD-HD outperforms both FD-IP and
HD since it assumes perfect knowledge at the source about the
instantaneous relay transmit power (as FD-IP) and can  work  in either 
FD or HD mode, depending on the system parameters.  As far as our
proposed technique is concerned, OP always outperforms HD and
achieves higher rates than FD-IP for $\Psav<-10$\,dBW. Furthermore, OP gets
very close to FD-HD, especially for $\Psav>P_1$.

Such performance of the OP scheme is achieved for the source and relay
transmit power levels and for the phase durations 
depicted in Figures~\ref{fig:figure135_slot_a}
and~\ref{fig:figure135_slot_b}, respectively.  Interestingly, for
$\Psav<P_1$, the time durations of the two communication
phases remain constant. With regard to the transmit power, for
$\Psav < P_0$, the source transmits in phase B and is silent in phase A
while the relay only transmits in phase A at its maximum power.  For
$P_0 \le \Psav < P_1$, the source always transmits (even if at
different power levels), while the relay only receives in phase B and
transmits at its maximum power in phase A. For $P_1 \le \Psav < P_2$, both
source and relay transmit but the duration of the two phases varies, with
$t_A\to 0$ as $\Psav\to P_2$. Finally, for $\Psav \ge P_2$,  both source and
relay transmit at their average power level.

Figure~\ref{fig:figure110} refers to the same scenario as that considered
in Figure~\ref{fig:figure135}, but with the self-interference attenuation
factor, $\beta$,   set to -110\,dB. In this case,
$\beta_0 = \beta/N_0 \approx 41$ dB and the results obtained for
$v < \beta_0$ apply.  Moreover,  we have: $P_0=1$\,dBW, $P_2=21$\,dBW,
$P_3=-9.9$\,dBW, and $P_4=-0.7$\,dBW, while the threshold $P_1$ is negative
(hence, it is not shown).  The figure highlights two operational
regions for $\Psav \ge P_0$ (namely, $P_0 \le\Psav\le P_2$ and
$\Psav \ge P_2$), and three operational regions for $\Psav < P_0$
(i.e., $\Psav < P_3$, $P_3 \le\Psav\le P_4$ and $P_4 < \Psav <
P_0$). In this case too, OP outperforms FD-IP (except for high values
of $\Psav$) and performs very close to FD-HD.  By looking at 
Figure~\ref{fig:figure110_slot_a}, which depicts the corresponding power levels used
at source and relay, we note that in phase B the relay
is always silent. In phase A, instead, the relay transmits at its
maximum power (namely, -7\,dBW) when $\Psav \le P_3$, and it slowly decreases its power to
$\Prav$ as $\Psav$ approaches $P_2$. With regard to the source, in
phase B it 
always transmits for $\Psav<P_2$, although at different  power levels depending on
$\Psav$. On the contrary,  in phase A it is silent for
$\Psav<P_4$, and it always transmits for larger values of $\Psav$. These results match the
values of the phase durations  depicted in
Figure~\ref{fig:figure110_slot_b}: now, the region where the 
phase durations are constant is limited to $\Psav < P_3$,
while, as $\Psav$ approaches $P_2$, $t_A\to 1$ and
$t_B\to 0$.

Figure~\ref{fig:figure3} highlights the impact of  self-interference on
the network performance. Indeed, the plot shows the rate versus $\Psav$, achieved by OP
and its counterpart FD-HD, as $\beta$ varies. For sake of
completeness, also the results for FD-Ideal and HD (which do not
depend on $\beta$) are shown. For $\beta=-120$\,dB (i.e.,
$v<\beta_0$), the system is affected by a substantial
self-interference at the relay, and OP performs as  HD for low-medium
values of $\Psav$. 
As $\beta$ decreases (i.e., the effect of self-interference is
smaller), 
the OP performance becomes closer to that of FD-HD and FD-Ideal; in
particular, for $\beta=-140$\,dB, the gap between OP and FD-HD reduces
to about 1\,dB.

In Figure~\ref{fig:figure_pP_3}, we study a different scenario
where $\beta$ is fixed to -130\,dB, 
$\Prav$ and $\Psav$ vary, and $\Prmax=\Prav+3$\,dB. Since now 
$\Psav$, $\Prav$ and $\Prmax$ can all grow very
large, the gap between FD-Ideal  and all other schemes becomes much
more evident. However, OP closely matches FD-HD and significantly
outperforms HD. Interestingly, FD-IP provides a lower rate than HD as 
the transmit power at source and relay increases. This is because 
FD-IP cannot exploit the HD mode; thus, when $\Prav$ is large
and the impact of self-interference becomes severe, there is no match 
with the other schemes. 

Finally, Figure~\ref{fig:figure_pP_40} addresses a similar
scenario to the one above, but  $\Prmax$ is now fixed to 10\,dBW.
We observe that, as $\Prav$ grows, the rate provided by all schemes
increases. However, when $\Prav$ approaches $\Prmax$, the relay is
constrained to transmit, i.e., to work in FD, for an increasingly
longer time. For $\Prav=\Prmax$, the relay always transmits at a power
level equal to $\Prav=\Prmax$. 
Also, the rate
provided by the HD scheme drops to 0 while FD-IP and FD-HD provide the
same performance; indeed, the latter cannot exploit anymore the
advantages of HD. For the same reason, the OP scheme experiences a
rate decrease. 
These results clearly suggest that significantly better performance
can be achieved when $\Prav$ is not too close to $\Prmax$.

\section{Extension to finite $\Psmax$\label{sec:extension}}
The analysis performed in Sections~\ref{sec:allocation}
and~\ref{sec:allocation2} as well as the numerical results reported
in Section~\ref{sec:results} have been obtained by assuming $\Psmax$
to be very large.  By relaxing  this assumption, the transmission power
at the source can be written as (see Appendix~\ref{app:A}): 
\begin{equation}
P(p) = \min\left\{\frac{\beta}{|h_1|^2}[\omega-p]^+,\Psmax\right\}\,.
\end{equation}
For simplicity, we define $\Pcsmax = \frac{|h_1|^2}{\beta}\Psmax$ so
that $P(p)$ can be more conveniently written as
$P(p)=\frac{\beta}{|h_1|^2}\Pc(p)$ with 
$\Pc(p) = \min\left\{[\omega-p]^+,\Pcsmax\right\}$.
\begin{figure}[t]
  \centerline{\resizebox{0.9\columnwidth}{!}{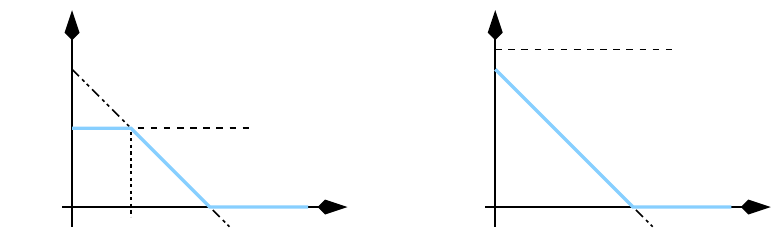}}
  \caption{Transmission power at the source, $\Pc(p)$, for $\Pcsmax\le \omega$ (left) and $\Pcsmax > \omega$ (right).}
  \label{fig:extension}
\end{figure}
The function $\Pc(p)$ is plotted in Figure~\ref{fig:extension} (blue line) for the
cases $\Pcsmax\le \omega$ (left) and $\Pcsmax > \omega$ (right).

The following cases can occur:
\begin{itemize}
\item $\Pcsmax > \omega$, then $\Pc(p) = [\omega-p]^+$. This leads to
  a situation similar to that considered in
  Sections~\ref{sec:allocation} and~\ref{sec:allocation2}. Indeed,
  \begin{itemize}
  \item in Section~\ref{sec:allocation}, by imposing
    $\omega \ge \Prmax$ in constraints (c), (d), and (e) of problem
    {\bf P1}, we obtained $\omega = \Psav+\Prav$. It follows that $\Pcsav$ should lie in the range
    $[\Prmax-\Prav,\Pcsmax-\Prav]$ where the results reported in
    Table~\ref{tab:1} hold;
   \item in Section~\ref{sec:allocation2}, we considered the case
  $\omega < \Prmax$.  Since $\Pcsmax > \omega$, we
  obtain $\omega < \min\{\Prmax,\Pcsmax\}$.  If
  $\Pcsmax>\Prmax$ the results shown in Table~\ref{tab:2} hold,
  otherwise they need to be recomputed by simply considering $\omega$ ranging in $[0,\Pcsmax)$.
   \end{itemize}
 \item $\Pcsmax \le \omega$, which is a more challenging scenario to
   analyse. Indeed, in such a situation function $\Pc(p)$, with 
   $p\in[0,\Prmax]$, takes values in up to three linear regions, depending on the
   value of $\Prmax$. Specifically,
   \begin{itemize}
   \item if $\Prmax<\omega - \Pcsmax$, we have $\Pc(p)=\Pcsmax$. Then
     the integral in~\eqref{eq:Ps-constraints} holds only if
     $\Pcsav=\Pcsmax$. This corresponds to the case where the source always
     transmits at its maximum power, regardless what the
     relay does;
   \item if $\omega - \Pcsmax \le \Prmax<\omega$, $\Pc(p)$ takes
     values in  two linear regions, i.e., $\Pc(p) = \Pcsmax$ if
     $p\in[0,\omega-\Pcsmax)$, and $\Pc(p) = \omega-p$ if
     $p\in[\omega-\Pcsmax,\Prmax)$. In order to maximize the rate $R$
     over the distribution $f(p)$, we then need to split it in two
     parts as done in Section~\ref{sec:allocation2} and the same
     analysis therein applies;
   \item if $\Prmax \ge \omega$ , $\Pc(p)$ is composed of three linear
     regions, i.e., $\Pc(p) = \Pcsmax$ if $p\in[0,\omega-\Pcsmax)$,
     $\Pc(p) = \omega-p$ if $p\in[\omega-\Pcsmax,\omega)$, and
     $\Pc(p)=0$ otherwise. In this case, for any given $\omega$, the rate
     maximization problem can be solved by splitting $f(p)$ in three
     distributions having support in $[0,\omega-\Pcsmax)$,
     $[\omega-\Pcsmax,\omega)$, and $[\omega,\Prmax]$, and having
     masses $F_1(\omega)$, $F_2(\omega)$, and
     $1-F_1(\omega)-F_2(\omega)$, respectively.  The rate
     maximization can be performed following a procedure similar to
     that used in Section~\ref{sec:allocation2}, although  in this case,
      we need to consider a three-dimensional (instead of a bi-dimensional) region $\Omega$, with
     coordinates $(\omega,F_1,F_2)$. Such maximization is quite 
     cumbersome if performed analytically, but quite easy to solve numerically.
   \end{itemize}
   \end{itemize}

\section{Conclusions\label{sec:conclusions}}
We investigated the maximum achievable rate in dual-hop
decode-and-forward networks where the relay can operate in full-duplex mode.
Unlike existing work, in our scenario the source must be aware only of
the distribution of the transmit power at the relay; under this assumption, we derived
the  allocation of the transmit power at the source and 
relay that maximize the data rate. Such distribution turned out 
to be discrete and composed of either one or two delta
functions. This finding  allowed us to identify the optimal
network communication strategy, which, in general, is given by a two-phase
time division scheme.  

Our numerical results highlight 
the advantage of being able to gauge full-duplex and  half-duplex 
at the relay, depending on the channel gains and the amount of
self-interference affecting the system. They also underline  the
excellent performance of the  proposed scheme, even when compared to strategies
that assume the source to be aware of the instantaneous transmit power
at the relay. 

\bibliographystyle{IEEEtran}
\bibliography{ref}

%%%%%%%%%%%%%%%%%%%%%%%%%%%%%%%%%%%%%%%%%%%%%%%%%%%%%%%%%%%%%%%%%%%%%%%%%%%%%%%%%%%%%%%%%%%%%%%%%%%%%%%%%%%%%%%%%%%%%%%%%%%%%%%%% 
\appendices

\section{Maximizing $R_1$ for a given $f(p)$}\label{app:A}
The optimization problem at hand is as follows: 
\begin{eqnarray}
&&\max \int_{0}^{\Prmax} f(p) \log \left(1+\frac{P(p)|h_1|^2}{N_0+\beta p}\right) \dd p \non
&s.t.& \non
& (a)& \int_{0}^{\Prmax} f(p) P(p)\dd p = \Psav \non
&(b) & 0 \leq P(p)\leq \Psmax \nonumber
  \label{eq:maxR1_z}
\end{eqnarray}
with $f(p)$ to be considered as a fixed arbitrary distribution.

We can solve the problem by  writing 
Lagrange's equation and leveraging the well-known Karush-Kuhn-Tucker (KKT) conditions.
We define the Lagrangian as:
\begin{eqnarray}
\Lc(P)&=& f(p) \log \left(1+\frac{|h_1|^2}{N_0+\beta p} P(p)\right ) -\lambda\left(f(p) P(p)\dd p-\Psav\right )\non 
&&-\mu_1(p) (P(p)-\Psmax) +\mu_2(p)  P(p) 
\end{eqnarray}
where $\mu_1(p), \mu_2(p) \geq 0$ and $\lambda$  are the KKT multipliers. 
Writing the KKT conditions, we obtain:
\begin{eqnarray}
\label{eq:KKT-A-1}
 \frac{|h_1|^2}{N_0+\beta p} \cdot\frac{f(p)}{1+\frac{P(p)|h_1|^2}{N_0+\beta p}}-\lambda f(p) -\mu_1(p) +\mu_2(p) &=&0\\
\mu_1(p) (P(p)-\Psmax) &=&0\\
\mu_2(p)  P(p) &=&0                    
\end{eqnarray}
along with (a) and (b) that must still hold. It can be easily verified that
the above system is satisfied when 
$\mu_1(p)=\mu_2(p)=0$, for which \eqref{eq:KKT-A-1} reduces to:
\begin{equation}
f(p) \left[\frac{|h_1|^2}{N_0+\beta p}\cdot \frac{1}{1+\frac{P(p)|h_1|^2}{N_0+\beta p}}-\lambda \right ]=0 \,.
\end{equation}
Excluding the trivial case where $f(p) =0$ $ \forall p$ and taking
into account the constraint $0\le P(p)\le \Psmax$, we get the following
expression for the optimal $P(p)$:
\begin{equation}
P(p)=\min \left\{\left[\frac{1}{\lambda}+\frac{N_0+\beta p}{|h_1|^2}\right]^+,\Psmax\right\} =\min\left\{ \frac{\beta[\omega-p]^+}{|h_1|^2},\Psmax\right\}
\label{eq:Ps-opt}
\end{equation}
where we defined
\[ \omega=\frac{|h_1|^2}{\beta\lambda}+\frac{N_0}{\beta}\]
which, in order to provide feasible solutions,  
must  satisfy (a). 

\section{\label{app:lemma1}}

\begin{lemma} \label{lemma:1}
  Let $\phi(p)$ be a continuous concave function in $p\in [a,b]$ and $f(p)$ be a probability distribution
  with support in $p\in [a,b]$ and average $\int_a^b p f(p) \dd p = m$. Then
  \[ p^*\phi(a)+(1-p^*)\phi(b) \le \int_a^b f(p)\phi(p) \dd p \le \phi(m) \]
  where $p^* = \frac{b-m}{b-a}$.
\end{lemma}
\begin{IEEEproof}
  The upper bound is obtained by applying Jensen inequality and holds with equality when
  $f(p)=\delta(p-m)$.
  
  With regard to the lower bound, being $\phi(p)$ concave in $p\in [a,b]$, we can write
  \[ \phi(p) \ge \frac{\phi(b)-\phi(a)}{b-a}(p-a) +\phi(a)\]
  Therefore, 
  \begin{eqnarray}
    \int_a^b f(p)\phi(p) \dd p 
    &\ge&  \int_a^b f(p)\left[\frac{\phi(b)-\phi(a)}{b-a}(p-a)+\phi(a)\right]\dd p \non
%    &=&   \frac{\phi(b)-\phi(a)}{b-a}\int_a^b f(p)(p-a)\dd p +\phi(a) \non
%    &=&   \frac{\phi(b)-\phi(a)}{b-a}(m-a)+\phi(a) \non
    &=&   \frac{b-m}{b-a}\phi(a)+\left(1-\frac{b-m}{b-a}\right)\phi(b) \,.
  \end{eqnarray}
  The lower bound holds with equality when
  \[
  f(p)=\frac{b-m}{b-a}\delta(p-a)+\left(1-\frac{b-m}{b-a}\right)\delta(p-b) \,.\]
\end{IEEEproof}

\section{Proof of Theorem~\ref{th:two_delta}\label{app:two_delta}}
  The calculus of variations problem in~\eqref{eq:theorem_problem} can be solved by using the Euler-Lagrange formula.
  To do so we first define the Lagrangian
  \[ L(z,f(p)) = f(p) \phi(p) +\lambda_1 f(p)\psi(p)+\lambda_2 p f(p) + \lambda_3 f(p) -\mu(p)f(p)\]
  where the first term represents the functional to be minimized. The second,
  third and fourth terms represent the constraints (a), (b), and (c) with
  associated Lagrange multipliers $\lambda_1$, $\lambda_2$ and $\lambda_3$,
  respectively. As far as the last term is concerned, we first observe that constraint (d) can be
  rewritten as $-f(p)\le 0$. In order to include (d) in the Lagrangian, we need
  to add a Lagrange multiplier for every $p\in[a,b]$. This can be done by
  introducing the multiplier $\mu(p)\ge 0$.

  Next, we apply the Euler-Lagrange formula and we write the Karush-Kuhn-Tucker
  conditions associated with the problem. Specifically, we get
  \[ \frac{\partial L}{\partial f} = 0 \Rightarrow  \mu(p) = \phi(p) +\lambda_1\psi(p)+\lambda_2 p + \lambda_3 \]
  subject to the conditions (a), (b), (c), (d), $\mu(p)\ge 0$, and $\mu(p)f(p)=0$.

  Now the key observation is that $\mu(p) = \phi(p)
  +\lambda_1\psi(p)+\lambda_2 p + \lambda_3$ identifies a family of
  continuous functions driven by the parameters $\lambda_1$, $\lambda_2$ and
  $\lambda_3$. Such parameters need to be properly chosen in order to
  have $\mu(p)\ge 0, \forall p\in[a,b]$.  If $\mu(p)$ is strictly
  positive in $[a,b]$ (i.e., $\mu(p)>0, \forall p\in[a,b]$), then the
  condition $\mu(p)f(p)=0$ implies $f(p)=0,\forall p\in[a,b]$, which is
  not a valid solution.  Moreover, $\phi(p)$ and $\psi(p)$ are not
  constant functions, therefore it is not possible to find values of
  the Lagrange multipliers such that $\mu(p)=0$, in a subset of
  $[a,b]$ having non-zero measure.
  The only option is to allow $\mu(p)>0$ for all $p\in[a,b]$, 
  except for a discrete set of points $p_i\in [a,b]$ for which
  $\mu(p_i)=0$. This observation hints that the solution of the
  problem must be found in the set of discrete distributions. In practice, every
  solution $p_i$ of $\mu(p)=0$ is associated with a mass of
  probability, $\pi_i$, located at $p_i$.

  The number of solutions of $\mu(p)=0$ can vary depending on the
  values of $\lambda_1$, $\lambda_2$, $\lambda_3$, $\gamma_1$, and
  $\gamma_2$. In general such a number can be computed by analyzing
  the first derivative of $\mu(p)$, i.e., 
  \begin{equation}\label{eq:mu'}
    \mu'(p) = \frac{k_1p^2+k_2p+k_3}{(1+\gamma_1 p)(1+\gamma_2 p)}
  \end{equation}
  where $k_1,k_2,k_3$ depend on $\lambda_1$, $\lambda_2$, $\lambda_3$, 
    $\gamma_1$, $\gamma_2$.

  The numerator of~\eqref{eq:mu'} is a polynomial in $p$ of degree 2
  and thus has up to two solutions for $p$ in $[a,b]$, which
  correspond to local minima or maxima of $\mu(p)$.

  Let $f^\star(p)$ be the minimizer
  of~\eqref{eq:theorem_problem}. Then several cases are possible:
  \begin{itemize}
  \item $\mu(p)$ has a single solution $p_1\in [a,b]$ which does not correspond to local minima or maxima. Then $p_1=a$ or $p_1=b$.
    This implies $f^\star(p) = \pi_1\delta(p-a)$ (or, $f^\star(p) = \pi_1\delta(p-b)$) which, however has only one degree of freedom (i.e., the
    value of $\pi_1$) and thus, in general, cannot satisfy 
    constraints (a), (b), and (c) of~\eqref{eq:theorem_problem} all together;
  \item $\mu(p)$ has a single solution $p_1\in [a,b]$ which
    corresponds to a local minimum. Thus
    $f^\star(p) = \pi_1\delta(p-p_1)$. However, this solution is not
    feasible since it has only two degrees of freedom (i.e., $p_1$ and
    $\pi_1$) and therefore, in general, cannot satisfy the three constraints (a),
    (b), and (c) of~\eqref{eq:theorem_problem} at the same time;
  \item  $\mu(p)$ has two solutions $p_1,p_2\in [a,b]$ none of which corresponds to a local minimum. Thus $p_1=a$ and $p_2=b$, and $f^\star(p) =
    \pi_1\delta(p-a)+\pi_2\delta(p-b)$.  Again, in general, this
    solution is not feasible since it has only two degrees of freedom
    ($\pi_1$ and $\pi_2$) and therefore cannot meet (a), (b), and (c)
    at the same time;
  \item  $\mu(p)$ has two solutions $p_1,p_2\in [a,b]$ one of which is a local minimum. Then two cases are possible, i.e., $\{p_1=a, p_2>a\}$ or $\{p_1<b,
    p_2=b\}$) and the minimizer $f^\star(p)$ takes the expression $f^\star(p) =
    \pi_1\delta(p-a)+\pi_2\delta(p-p_2)$ or $f^\star(p) = \pi_1\delta(p-p_1)+\pi_2\delta(p-b)$.
    This solution is feasible since it has three degrees of freedom represented by
    $\{\pi_1, \pi_2, p_1\}$ or $\{\pi_1, \pi_2, p_2\}$ that can be determined by imposing the constraints (a), (b), and (c).
    The constants $\gamma_1$, and $\gamma_2$ determine which of the two expressions
    in~\eqref{eq:theorem_sol} is the minimizer.  This is shown in Section~\ref{app_two_delta_1}.  
  \end{itemize}
  Since $\mu(p)$ cannot have more than two distinct solutions in
  $[a,b]$, (as it can be observed from the fact that $\mu'(p)$ has at
  most two solutions), we conclude that the minimizer
  of~\eqref{eq:theorem_problem} is given by~\eqref{eq:theorem_sol}.

  \subsection{Selecting the minimizer expression\label{app_two_delta_1}}
As shown above, the minimizer can assume one of the two possible
expressions reported in~\eqref{eq:theorem_sol}. Here we  show that the
choice of the minimizer depends on the parameters $\gamma_1$ and
$\gamma_2$.  To do so, we first observe that the family of
distributions
\begin{equation}
  \label{eq:general_solution}
  f^\star(p,x,y) = \pi(x,y) \delta(p-x) + [1-\pi(x,y)]\delta(p-y)\,,
\end{equation}
where
\[ \pi(x,y) = \frac{y-m}{y-x}>0\,,\]
with $m \le y \le b$ and $a\le x \le m$, encompasses both expressions
in~\eqref{eq:theorem_sol}.  Specifically, the expressions reported
in~\eqref{eq:theorem_sol} are given by $f^\star(p,a,p_2)$ and 
$f^\star(p,p_1,b)$, respectively.

For the family of distributions in~\eqref{eq:general_solution}, 
constraint (c) in~\eqref{eq:theorem_problem} can be rewritten as
\begin{eqnarray}
  F(x,y) &=& \int_a^b f^\star(p,x,y) \psi(p)\dd p\non
  &=& \pi(x,y) \psi(x)+ [1-\pi(x,y)] \psi(y)\non
  &=& c \,.
\end{eqnarray}
Similarly, the cost function, $\int_a^bf(p)\phi(p) \dd p$,  can be written as
\[  G(x,y) = \pi(x,y) \phi(x) + [1-\pi(x,y)]\phi(y) \,.\]
Observe that, since $\psi(p) = \phi(p)+\eta(p)$, we have $F(x,y) =
G(x,y)+H(x,y)$ where
\[ H(x,y) = \pi(x,y) \eta(x) + [1-\pi(x,y)]\eta(y) \,.\]
In the following, for the sake of notation simplicity, we drop the
argument of the functions when not needed.
We now make the following observations:
\begin{enumerate}
\item $F$ and $G$ are increasing functions of $x$ and decreasing
  functions of $y$.  Indeed, for any concave function $\rho(p)$,
  $p\in[a,b]$ (as $\phi(p)$, $\eta(p)$ and $\psi(p)$) the partial
  derivative of $\pi\rho(x) +(1-\pi)\rho(y)$ is given by
\[ \frac{\partial [\pi\rho(x) +(1-\pi)\rho(y)]}{\partial x} =   
\frac{\pi}{y-x}\left(\rho'(x)-\frac{\rho(y)-\rho(x)}{y-x}\right) > 0 \,.\]
The factor $\pi/(y-x)$ is clearly positive since $\pi$ is positive by
definition and $y > x$. Moreover, for a concave function, the
difference quotient $\frac{\rho(y)-\rho(x)}{y-x}$ is smaller than the
derivative of $\rho(p)$ computed in $x$.  Similarly, it is
straightforward to show that
\[ \frac{\partial [\pi\rho(x) +(1-\pi)\rho(y)]}{\partial y} < 0 \,;\]

\item the equation $F(x,y) = c$ is the implicit definition of the
function $y_c(x)$, $a\le x\le p_1$ and $p_2\le y\le b$ with derivative 
defined as
\[ y_c'(x) = \frac{\dd y_c(x)}{\dd x} = -\frac{\frac{\partial F}{\partial x}}{\frac{\partial F}{\partial y}} = -\frac{F_x}{F_y}\,, \]
where, for simplicity, we defined $F_x= \frac{\partial F}{\partial x}$
and $F_y= \frac{\partial F}{\partial y}$.  By the above arguments on
the partial derivatives of $F$, we conclude that $y_c'(x) >0$.
Similarly, the function $G(x,y) = t$ is the implicit definition of the
function $y_t(x)$ whose derivative $y_t'(x)$ is positive.
\item Given the constant $c$, a value for $t$ exists such that
  $y_c(x)$ and $y_t(x)$ have a common solution $(x^*,y^*)$. For
  example, if $t=G(a,p_2)$, the two curves share the point $(a,p_2)$
  where $p_2=y_c(a)$.
\end{enumerate}

Now consider a value of $t$ such that the curves $y_c(x)$ and $y_t(x)$
intersect at point $P=(x^*, y^*)$, with $P\neq (a,p_2), P\neq (p_1,b)$.

If $y_c'(x)>y_t'(x)$ at $P$ then $t$ is not the global minimum of the cost function
in~\eqref{eq:theorem_problem}. Indeed, it exists
$\epsilon>0$ such that the curve $y_{t-\epsilon}(x)$ intersects
$y_{c}(x)$ at some point $P'=(x^*+\Delta_x,y^*+\Delta_{y})$ where the
cost function $G(x^*+\Delta_x,y^*+\Delta_{y})=t-\epsilon$ is clearly
lower than at $P$. Since this is true for any point $P=(x^*, y^*)$ we conclude 
that the minimizer is $f^\star(p,p_1,b)$ and that the minimum
is thus $G(p_1,b)$. By applying similar arguments, if $y_c'(x)<y_t'(x)$ at $P$
the minimizer is $f^\star(p,a,p_2)$ and the minimum is $G(a,p_2)$.

In order to compare the derivatives of $y_c'(x)$ and $y_t'(x)$ we use
the definitions of $F$, $G$, and $H$ and write the following
differential equations
\[  y_c'(x) = -\frac{F_x}{F_y} = -\frac{G_x+H_x}{G_y+H_y}\,,\hspace{4ex} y_{t}'(x) = -\frac{G_x}{G_y} \]
where $G_x$, $H_x$, $G_y$, $H_y$ are the partial derivatives of $G$
and $H$ w.r.t. $x$ and $y$, respectively.
By considering  $y_c'(x) \ge y_t'(x)$, we obtain
\[ -\frac{G_x+H_x}{G_y+H_y} \ge -\frac{G_x}{G_y} \Rightarrow  -\frac{G_x}{G_y}\le -\frac{H_x}{H_y} \,.\]
We also observe that
\[ \frac{\partial \pi}{\partial x} = \pi_x = \frac{\pi}{y-x}\,; \hspace{6ex} \frac{\partial \pi}{\partial y} = \pi_y = \frac{1-\pi}{y-x}\,. \]
Therefore, we have
\begin{eqnarray}
  G_x &=& \pi_x\phi(x) +\pi \phi'(x) -\pi_x\phi(y) = \pi \left[\phi'(x)-\frac{\phi(y)-\phi(x)}{y-x}\right] \non
  G_y &=& \pi_y\phi(x) -\pi_y\phi(y) +(1-\pi)\phi'(y) = (1-\pi)\left[\phi'(y)- \frac{\phi(y)-\phi(x)}{y-x}\right] \non
  H_x &=& \pi_x\eta(x) +\pi \eta'(x) -\pi_x\eta(y) = \pi \left[\eta'(x)-\frac{\eta(y)-\eta(x)}{y-x}\right] \non
  H_y &=& \pi_y\eta(x) -\pi_y\eta(y) +(1-\pi)\eta'(y) =
  (1-\pi)\left[\eta'(y)- \frac{\eta(y)-\eta(x)}{y-x}\right] \,.\non
\end{eqnarray}

Now observe that $\phi(p)=\log(1+\gamma_1 p)$ and $\eta(p)=\log(1+\gamma_2 p)$
are the same function (i.e., $\log(1+\gamma p)$), the former evaluated in
$\gamma=\gamma_1$ and the latter in $\gamma=\gamma_2$. Therefore, since $G$
depends on $\phi(p)$ and $H$ depends on $\eta(p)$, we can write
$-\frac{G_x}{G_y}=\zeta(\gamma_1)$ and $-\frac{H_x}{H_y}=\zeta(\gamma_2)$.

It is easy to show that $\zeta(\gamma)$ increases with $\gamma$;
indeed, by imposing $\zeta'(\gamma)\ge 0$, after some algebra and after simplifying positive factors, we obtain
\[\log\left( \frac{1+\gamma y}{1+\gamma x}\right)(2+\gamma y+\gamma x) \ge 2\gamma (y-x) \,.\]
The right hand side  (r.h.s.) of the previous inequality is positive and linear with $\gamma$,
$\gamma\ge 0$. The left hand side  (l.h.s.) is positive, convex and tangent to the r.h.s. at
$\gamma=0$. Therefore, the above inequality always holds and $\zeta(\gamma)$
increases with $\gamma$. We conclude that if $\gamma_1<\gamma_2$, we have
$-\frac{G_x}{G_y}\le -\frac{H_x}{H_y}$ and, thus, $y_c(x) > y_t(x)$. In such a
case the minimizer is $f^\star(p,p_1,b)$.  Similarly, when $\gamma_1>\gamma_2$,
the minimizer is $f^\star(p,a,p_2)$.

\section{Behavior of the curve $Q_1(\omega,F)=0$\label{app:varie}}
%\subsection{About the curve $Q_1(\omega,F)=0$}
The curve $Q_1(\omega,F)$ intersects the line $\omega = \Prav+\Pcsav$
at most in a single point.  To prove this, we substitute
$\omega = \Prav+\Pcsav$ in the expression for $Q_1(\omega,F)=0$, i.e.,
we compute 
$(R_1^{\rm min}=\widetilde{R}_2^{\rm max})|_{\omega =
  \Prav+\Pcsav}$. After some algebra and by setting
$a=\beta_0(\Prav +\Pcsav)$, we obtain: 
  \begin{equation} -\log \left( 1-\frac{\Pcsav\beta_0}{F(1+a)}\right) =F \log\left(1-\frac
      {\Pcsav\beta_0 v}{F(\beta_0+a v)}\right) + \log\left(
      1+\frac {a v}{\beta_0} \right) \,.\label{eq:Q1_one_intersection}
  \end{equation}
  Observe that the l.h.s of~\eqref{eq:Q1_one_intersection} is defined
  when $\frac{\Pcsav\beta_0}{F(1+a)}<1$, i.e., when
  $F\ge \frac{\Pcsav\beta_0}{1+a}$, which is always true since in
  $\Omega$ $F$ is larger that the value it achieves in $V_3$, i.e.,
  $\frac{\Pcsav\beta_0}{a}$.  Moreover, the l.h.s
  of~\eqref{eq:Q1_one_intersection} decreases with $F$ while the r.h.s
  of~\eqref{eq:Q1_one_intersection} increases with
  $F$; thus,~\eqref{eq:Q1_one_intersection} has at most one solution.

\section{Behavior of $R_1^{\rm min}$,$R_1^{\rm max}$, $R_2^{\rm min}$,$R_2^{\rm max}$\label{app:Rminmax_omegaF}}

\subsection{$R_1^{\rm min}$}
We first observe that $R_1^{\rm min}$ can be rewritten as
\[ R_1^{\rm min}  = F\log\left(1+ \frac{\Pcsav\beta_0}{F(1+\omega \beta_0) -\Pcsav\beta_0}\right)\,. \]
From the above expression, we immediately observe that $ R_1^{\rm min}$ decreases with $\omega$.
We then compute the partial derivative of $R_1^{\rm min}$ w.r.t. $F$. We have
\[\frac{\partial}{\partial F} R_1^{\rm min} = \log\left(1+\frac{a}{F-a}\right)-\frac{a}{F-a}\]
where $a=\Pcsav\beta_0/(1+\omega \beta_0)$. Then $\frac{\partial}{\partial F} R_1^{\rm min} \le 0$ implies
\[ \log\left(1+\frac{a}{F-a}\right) \le  \frac{a}{F-a} \,.\]
Since $\log(1+y)\le y$, for $y>-1$, we conclude that $R_1^{\rm min}$ decreases with $F$.

\subsection{$R_1^{\rm max}$}
$R_1^{\rm max}$ does not depend on $F$; moreover, 
\[\frac{\partial}{\partial \omega} R_1^{\rm max} = -\Pcsav \frac{(1+\beta_0\omega)\log(1+\beta_0\omega)-\beta_0\omega}{\omega^2(1+\beta_0\omega)}\,.\]
Therefore, $\frac{\partial}{\partial \omega} R_1^{\rm max}\le 0$
implies 
$(1+\beta_0\omega)\log(1+\beta_0\omega)>\beta_0\omega$, which 
 is always true since $\beta_0 \omega\ge 0$ and $y \le (1+y) \log (1+y)$ for $y \ge 0$.
Hence, $R_1^{\rm max}$ decreases with $\omega$.

\subsection{$R_2^{\rm min}$}
We now consider the expression for $\widetilde{R}_2^{\rm min}$ reported in~\eqref{eq:tilde_R2_min}, which can be rewritten as
\[\widetilde{R}_2^{\rm min} = (1-F)\log\left(1+\frac{a}{1-F}\right)+c\]
where $a>0$ and $c$ do not depend on $F$. Now
\[ \frac{\partial}{\partial F}\widetilde{R}_2^{\rm min} = 
\frac{\frac{a}{1-F}}{1+\frac{a}{1-F}}-\log\left(1+\frac{a}{1-F}\right)\,.\]
Thus, $\frac{\partial}{\partial F}\widetilde{R}_2^{\rm min}\le 0$ implies
\[ \frac{a}{1-F} \le
    \left(1+\frac{a}{1-F}\right)\log\left(1+\frac{a}{1-F}\right) \,.\]
  Since $y\le (1+y)\log(1+y)$ for $y>0$, the above statement is
  true; hence, $\widetilde{R}_2^{\rm min}$ decreases with $F$.

We now consider the partial derivative w.r.t. $\omega$:
\[ \frac{\partial}{\partial \omega}\widetilde{R}_2^{\rm min} =
  \frac{\Pcsav}{\omega^2} \log(1+\omega v)+v\frac{\left(F-\frac{\Pcsav}{\omega}\right)}{1+\omega v}
  -\frac{v F}{1+\omega v+v\frac{\Prav+\Pcsav-\omega}{1-F}} \,.
\]
We observe that the arguments of the logarithms are always
positive. Moreover, the last term is negative since $F>0$ and
$\Prav+\Pcsav\ge\omega$, while the second term is
positive since $F\ge \frac{\Pcsav}{\omega}$.  We now make the
following key observation about the last term:
\[ \frac{v F}{1+\omega v+v\frac{\Prav+\Pcsav-\omega}{1-F}}
  \le \frac{v F}{1+\omega v} \,.\]
Thus,
 \begin{eqnarray}
   \frac{\partial}{\partial \omega}\widetilde{R}_2^{\rm min}
   &\ge&  \frac{\Pcsav}{\omega^2} \log(1+\omega v)
         +v\frac{F-\frac{\Pcsav}{\omega}}{1+\omega v}
         -\frac{v F}{1+\omega v} \non
   &=&  \frac{\Pcsav}{\omega^2} \log(1+\omega v)
       -\frac{\Pcsav v}{\omega(1+\omega v)}
 \end{eqnarray}
 which is positive since 
 $(1+\omega v)\log(1+\omega v) \ge \omega v$. It follows that $\widetilde{R}_2^{\rm min}$ increases with $\omega$.

\subsection{$R_2^{\rm max}$}
By deriving $\widetilde{R}_2^{\rm max}$ in~\eqref{eq:tilde_R2_min} w.r.t. $\omega$, we get
\[ \frac{\partial}{\partial \omega}\widetilde{R}_2^{\rm max} =
  \frac{(F\Prav-\omega F+\Pcsav) v^2}{(1-F)\left(1+v\frac{\Prav+\Pcsav-\omega F}{1-F}\right)
    \left(1+v\left(\omega-\frac{\Pcsav}{F}\right)\right)} \,.\]
Since the arguments of the logarithms are positive, so is the denominator. Therefore, by imposing
$\frac{\partial}{\partial \omega}\widetilde{R}_2^{\rm max}\ge 0$ and
solving for $\omega$, we get
$\omega \le \Prav + \frac{\Pcsav}{F}$.  Since $F<1$, then
$\Prav + \frac{\Pcsav}{F} \ge \Prav+\Pcsav$. Since in
the region $\Omega$ we have $\omega \le \Prav+\Pcsav$, we conclude that $\widetilde{R}_2^{\rm max}$ increases with $\omega$.

\section{Maximizing the rate over subregion $\Omega_3$\label{app:region3}}
\begin{itemize}
\item For $v \ge \beta_0$, according to Theorem~\ref{th:two_delta}, the function $g(p)$ for which {\bf P4} is maximized is given by 
\begin{equation}
  g(p) = \frac{\frac{\Pcsav}{F}}{\omega-p_1}\delta(p-p_1)
  +\left(1-\frac{\frac{\Pcsav}{F}}{\omega-p_1}\right)\delta(p-\omega) \,.
\end{equation}
Substituting this expression in~\eqref{eq:problem_region3}, we can
rewrite the problem as: 
 \begin{eqnarray}
   R_{\Omega_3}&=&\max_{(\omega,F)\in \Omega_3}\frac{\Pcsav}{\omega-p_1}\log\left(\frac{1+\beta_0\omega}{1+\beta_0p_1}\right) \label{eq:Romega3_2}\\
  && {\rm s.t.} \non
   &&  \frac{\Pcsav}{\omega-p_1}\log\left[\frac{(1+\beta_0 p_1)(1+v p_1)}{(1+\beta_0 \omega)(1+v \omega)}\right] =
      C_1(\omega,F)\label{eq:constraint_C1(F)} \\
  && C_1(\omega,F) = F
  C(\omega,F)-F\log\left[(1+\beta_0\omega)(1+v\omega)\right] \,.\label{eq:C1(F)}
 \end{eqnarray}
 We first observe that the argument of~\eqref{eq:Romega3_2} decreases
 with $p_1$. Indeed, 
 \begin{eqnarray}
   \frac{\partial}{\partial p_1}\frac{\log\left(\frac{1+\beta_0\omega}{1+\beta_0 p_1}\right)}{\omega-p_1}
   &=& \frac{1}{(\omega-p_1)^2}\left[\log\left(\frac{1+\beta_0\omega}{1+\beta_0 p_1}\right) -\frac{\beta_0(\omega-p_1)}{1+\beta_0 p_1}\right] \non
   &\le & \frac{1}{(\omega-p_1)^2}\left[\frac{1+\beta_0\omega}{1+\beta_0 p_1}-1 -\frac{\beta_0(\omega-p_1)}{1+\beta_0 p_1}\right] \non
          &=&0
   \end{eqnarray}
   where we used the bound $\log y \le y-1$.
   %If we neglect the constraint in~\eqref{eq:constraint_C1(F)},
   %the variable $p_1$ does not depend on $\omega$ and $F$ and can assume any value in $[0,\omega-\Pcsav/F]$.  
   %we can bound the rate $R_{\Omega_3}$ as follows:
   %\[ R_{\Omega_3} = \max_{(\omega,F)\in \Omega_3}\frac{\Pcsav}{\omega-p_1}\log\left(\frac{1+\beta_0\omega}{1+\beta_0p_1}\right) \le 
   %   \max_{(\omega,F)\in \Omega_3}\frac{\Pcsav}{\omega-p_1}\log\left(\frac{1+\beta_0\omega}{1+\beta_0p_1}\right) 

   Also, the
 l.h.s. of~\eqref{eq:constraint_C1(F)} increases with $p_1$. This can
 be easily seen considering that, if
 we let  $\zeta(p) = \frac{(1+\beta_0 p)(1+v p)}{(1+\beta_0 \omega)(1+v \omega)}>0$, we
 can write
 \begin{equation}
   \frac{\partial}{\partial p}\frac{\log\zeta(p)}{\omega-p} = \frac{\zeta'(p)}{\zeta(p)(\omega-p)}+\frac{\log \zeta(p)}{(\omega-p)^2} = \frac{\zeta'(p)(\omega-p)+\zeta(p)\log\zeta(p)}{\zeta(p)(\omega-p)^2}\,.\label{eq:zeta'}
   \end{equation}
   Note that the denominator of the above equation is positive. Now
   $\zeta(p) = \zeta_1(p)\zeta_2(p)$ where
   $\zeta_1(p) = \frac{1+\beta_0 p}{1+\beta_0 \omega}$ and
   $\zeta_2(p) = \frac{1+v p}{1+v\omega}$.  It follows that the
   numerator of~\eqref{eq:zeta'} can be rewritten as
 \begin{eqnarray}
   \zeta'(p)(\omega-p)+\zeta(p)\log\zeta(p)
   &=& \zeta'(p)(\omega-p)+\zeta_2(p)\zeta_1(p)\log\zeta_1(p)+\zeta_1(p)\zeta_2(p)\log\zeta_2(p) \non
   &\ge& \zeta'(p)(\omega-p)+\zeta_2(p)[\zeta_1(p)-1]+\zeta_1(p)[\zeta_2(p)-1]\non
   &=& \zeta'(p)(\omega-p)+2\zeta(p)-\zeta_1(p)-\zeta_2(p)\non
   &=& 0
 \end{eqnarray}
 where we used the bound $y\log y \ge y-1$ which holds for any $y>0$.
 With regard to $C_1(\omega,F)$, we consider the curves $Q_2(\omega,F)=t$
  where $t < 0$ is a parameter. Note that such curves are located on
  the left of  $Q_2(\omega,F)=0$. 
By definition of $Q_2(\omega,F)$, we have
  \begin{eqnarray}
    Q_2(\omega,F) &=& \widetilde{R}_2^{\rm min}-R_1^{\rm max} \non
    &=&\left(F-\frac{\Pcsav}{\omega}\right)\log(1+v\omega)+(1-F)\log\left(1+v\frac{\Pcsav+\Prav-F\omega}{1-F}\right) \non
        &&\qquad -\frac{\Pcsav}{\omega}\log(1+\beta_0\omega) \non
    &=&\left(F-\frac{\Pcsav}{\omega}\right)\log\left[(1+\beta_0\omega)(1+v\omega)\right] 
        -F C(\omega,F) \non
    &=& t \,.
  \end{eqnarray}
  Therefore, the term $ C_{1,t}(\omega,F)$ defined in~\eqref{eq:C1(F)} can be written as
  \[ C_{1,t}(\omega,F) = -t
    -\frac{\Pcsav}{\omega}\log\left[(1+\beta_0\omega)(1+v\omega)\right]\,.\]
  The subscript $t$ indicates that we are restricting our analysis to 
  the curve $Q_2(\omega,F)=t$. Observe that the expression for
  $C_{1,t}(\omega,F)$ does not depend on $F$; thus, as $t$ increases,
  $C_{1,t}(\omega,F)$ decreases. Since the
  l.h.s. of~\eqref{eq:constraint_C1(F)} increases with $p_1$,  the
  value of $p_1$ for which the constraint is met decreases as $t$
  increases. Finally, since the argument of~\eqref{eq:Romega3_2}
  decreases with $p_1$, we conclude that the rate increases as $t$
  increases.

  Since the curves $Q_2(\omega,F)=t$ have positive derivative in the
  $(\omega,F)$ plane, and as $t$ increases they move to the right,
  then, for any fixed $\hat{\omega}$, the solution for $F$ of
  $Q_2(\hat{\omega},F)=t$ decreases as $t$ increases.
  It follows that the rate $R_{\Omega_3}$ is achieved on curve $B$--$D$ as well.
  Thus, $R=R_{\Omega_1}=R_{\Omega_3}$ and $R$ can be computed by solving
  $R=\max_{Q_1(\omega,F)=0}R_1^{\rm min}$.
\item   For $v  < \beta_0$, according to Theorem~\ref{th:two_delta},
  $g(p)$ of {\bf P4} is given by 
  \begin{equation}\label{eq:g(p)_omega3_2}
  g(p) = \frac{p_2-\omega+\frac{\Pcsav}{F}}{p_2}\delta(p)+\frac{\omega-\frac{\Pcsav}{F}}{p_2}\delta(p-p_2)\,.
\end{equation}
Replacing $g(p)$ in~\eqref{eq:problem_region3} with the above
expression, we get 
\begin{equation}\label{eq:R_Omega3_2}
  R_{\Omega_3} = \max_{(\omega,F)\in \Omega_3} F \log(1+\beta_0\omega)+ \frac{\Pcsav-F\omega}{p_2}\log(1+\beta_0 p_2)
\end{equation}
under the constraint
\begin{eqnarray} \label{eq:constraint_region3_2}\frac{\log\left[(1+\beta_0 p_2)(1+v p_2)\right]}{p_2}
%  &=& \frac{C(\omega,F)}{\omega-\frac{\Pcsav}{F}}\non
  &=& \frac{\log(1+\beta_0\omega)
    +\left(1-\frac{1}{F}\right)\log\left(1+v\frac{\Pcsav+\Prav-F\omega}{1-F}\right)}{\omega-\frac{\Pcsav}{F}} \,.
\end{eqnarray}
In this case, an analytic solution of the optimization problem cannot
be easily obtained since the involved terms do not exhibit the
monotonic behavior observed above. By solving the problem numerically, 
it turns out that the optimum is located in $A=(\omega_A,F_A)$ for
$\Pc_3 \le \Pcsav\le \Pc_4$, and in $V_2=(\Pcsav+\Prav,1)$ for $\Pc_4 \le \Pcsav\le \Pc_2$.
  \end{itemize}

\end{document}

%% file: frame.pdf_t
\begin{picture}(0,0)%
\includegraphics{frame.pdf}%
\end{picture}%
\setlength{\unitlength}{4144sp}%
\begingroup\makeatletter\ifx\SetFigFont\undefined%
\gdef\SetFigFont#1#2#3#4#5{%
  \reset@font\fontsize{#1}{#2pt}%
  \fontfamily{#3}\fontseries{#4}\fontshape{#5}%
  \selectfont}%
\fi\endgroup%
\begin{picture}(2679,1962)(259,-1597)
\put(1981,-1456){\makebox(0,0)[lb]{\smash{{\SetFigFont{8}{9.6}{\rmdefault}{\mddefault}{\updefault}{\color[rgb]{0,0,0}$p_B$}%
}}}}
\put(496,-601){\makebox(0,0)[lb]{\smash{{\SetFigFont{10}{12.0}{\rmdefault}{\mddefault}{\updefault}{\color[rgb]{0,0,0}$f(p)$}%
}}}}
\put(1486,-1546){\makebox(0,0)[lb]{\smash{{\SetFigFont{8}{9.6}{\rmdefault}{\mddefault}{\updefault}{\color[rgb]{0,0,0}$\Prav$}%
}}}}
\put(2476,-1546){\makebox(0,0)[lb]{\smash{{\SetFigFont{8}{9.6}{\rmdefault}{\mddefault}{\updefault}{\color[rgb]{0,0,0}$\Prmax$}%
}}}}
\put(2881,-196){\makebox(0,0)[lb]{\smash{{\SetFigFont{10}{12.0}{\rmdefault}{\mddefault}{\updefault}{\color[rgb]{0,0,0}$t$}%
}}}}
\put(2881,-1456){\makebox(0,0)[lb]{\smash{{\SetFigFont{10}{12.0}{\rmdefault}{\mddefault}{\updefault}{\color[rgb]{0,0,0}$p$}%
}}}}
\put(811,-1456){\makebox(0,0)[lb]{\smash{{\SetFigFont{8}{9.6}{\rmdefault}{\mddefault}{\updefault}{\color[rgb]{0,0,0}$p_A$}%
}}}}
\put(2026,254){\makebox(0,0)[lb]{\smash{{\SetFigFont{8}{9.6}{\rmdefault}{\mddefault}{\updefault}{\color[rgb]{0,0,0}$t_B$}%
}}}}
\put(811,-16){\makebox(0,0)[lb]{\smash{{\SetFigFont{8}{9.6}{\rmdefault}{\mddefault}{\updefault}{\color[rgb]{0,0,0}$P_A, p_A$}%
}}}}
\put(1846,-16){\makebox(0,0)[lb]{\smash{{\SetFigFont{8}{9.6}{\rmdefault}{\mddefault}{\updefault}{\color[rgb]{0,0,0}$P_B, p_B$}%
}}}}
\put(991,254){\makebox(0,0)[lb]{\smash{{\SetFigFont{8}{9.6}{\rmdefault}{\mddefault}{\updefault}{\color[rgb]{0,0,0}$t_A$}%
}}}}
\put(1846,-871){\makebox(0,0)[lb]{\smash{{\SetFigFont{8}{9.6}{\rmdefault}{\mddefault}{\updefault}{\color[rgb]{0,0,0}$t_B$}%
}}}}
\put(676,-1051){\makebox(0,0)[lb]{\smash{{\SetFigFont{8}{9.6}{\rmdefault}{\mddefault}{\updefault}{\color[rgb]{0,0,0}$t_A$}%
}}}}
\put(1486,-241){\makebox(0,0)[lb]{\smash{{\SetFigFont{8}{9.6}{\rmdefault}{\mddefault}{\updefault}{\color[rgb]{0,0,0}$1$}%
}}}}
\end{picture}%

%% file: triangle.pdf_t
\begin{picture}(0,0)%
\includegraphics{triangle.pdf}%
\end{picture}%
\setlength{\unitlength}{4144sp}%
\begingroup\makeatletter\ifx\SetFigFont\undefined%
\gdef\SetFigFont#1#2#3#4#5{%
  \reset@font\fontsize{#1}{#2pt}%
  \fontfamily{#3}\fontseries{#4}\fontshape{#5}%
  \selectfont}%
\fi\endgroup%
\begin{picture}(3987,3100)(301,-2384)
\put(361,164){\makebox(0,0)[lb]{\smash{{\SetFigFont{12}{14.4}{\rmdefault}{\mddefault}{\updefault}{\color[rgb]{0,0,0}1}%
}}}}
\put(316,569){\makebox(0,0)[lb]{\smash{{\SetFigFont{12}{14.4}{\rmdefault}{\mddefault}{\updefault}{\color[rgb]{0,0,0}$F$}%
}}}}
\put(4096,-2311){\makebox(0,0)[lb]{\smash{{\SetFigFont{12}{14.4}{\rmdefault}{\mddefault}{\updefault}{\color[rgb]{0,0,0}$\omega$}%
}}}}
\put(3643,141){\makebox(0,0)[lb]{\smash{{\SetFigFont{10}{12.0}{\rmdefault}{\mddefault}{\updefault}{\color[rgb]{0,0,0}$V_2$}%
}}}}
\put(1010,-1119){\makebox(0,0)[lb]{\smash{{\SetFigFont{10}{12.0}{\rmdefault}{\mddefault}{\updefault}{\color[rgb]{0,0,0}$V_1$}%
}}}}
\put(3623,-1655){\makebox(0,0)[lb]{\smash{{\SetFigFont{10}{12.0}{\rmdefault}{\mddefault}{\updefault}{\color[rgb]{0,0,0}$V_3$}%
}}}}
\put(1846,-1051){\makebox(0,0)[lb]{\smash{{\SetFigFont{10}{12.0}{\rmdefault}{\mddefault}{\updefault}{\color[rgb]{0,0,0}$\Omega_1$}%
}}}}
\put(2611,-826){\makebox(0,0)[lb]{\smash{{\SetFigFont{10}{12.0}{\rmdefault}{\mddefault}{\updefault}{\color[rgb]{0,0,0}$\Omega_3$}%
}}}}
\put(3106,-1186){\makebox(0,0)[lb]{\smash{{\SetFigFont{10}{12.0}{\rmdefault}{\mddefault}{\updefault}{\color[rgb]{0,0,0}$\Omega_2$}%
}}}}
\put(2435,-765){\rotatebox{295.0}{\makebox(0,0)[lb]{\smash{{\SetFigFont{6}{7.2}{\rmdefault}{\mddefault}{\updefault}{\color[rgb]{0,0,0}$Q_1(\omega,F)=0$}%
}}}}}
\put(2929,-1231){\rotatebox{70.0}{\makebox(0,0)[lb]{\smash{{\SetFigFont{6}{7.2}{\rmdefault}{\mddefault}{\updefault}{\color[rgb]{0,0,0}$Q_2(\omega,F)=0$}%
}}}}}
\put(3466,-2311){\makebox(0,0)[lb]{\smash{{\SetFigFont{10}{12.0}{\rmdefault}{\mddefault}{\updefault}{\color[rgb]{0,0,0}$\Pcsav+\Prav$}%
}}}}
\end{picture}%

%% file: triangle2.pdf_t
\begin{picture}(0,0)%
\includegraphics{triangle2.pdf}%
\end{picture}%
\setlength{\unitlength}{4144sp}%
\begingroup\makeatletter\ifx\SetFigFont\undefined%
\gdef\SetFigFont#1#2#3#4#5{%
  \reset@font\fontsize{#1}{#2pt}%
  \fontfamily{#3}\fontseries{#4}\fontshape{#5}%
  \selectfont}%
\fi\endgroup%
\begin{picture}(4181,3100)(118,-2452)
\put(1526,-1584){\makebox(0,0)[lb]{\smash{{\SetFigFont{6}{7.2}{\rmdefault}{\mddefault}{\updefault}{\color[rgb]{0,0,0}$\Omega_3$}%
}}}}
\put(362,-368){\makebox(0,0)[lb]{\smash{{\SetFigFont{6}{7.2}{\rmdefault}{\mddefault}{\updefault}{\color[rgb]{0,0,0}$V_1$}%
}}}}
\put(380,-1774){\makebox(0,0)[lb]{\smash{{\SetFigFont{6}{7.2}{\rmdefault}{\mddefault}{\updefault}{\color[rgb]{0,0,0}$V_1$}%
}}}}
\put(1760,362){\makebox(0,0)[lb]{\smash{{\SetFigFont{6}{7.2}{\rmdefault}{\mddefault}{\updefault}{\color[rgb]{0,0,0}$V_2$}%
}}}}
\put(1760,-1243){\makebox(0,0)[lb]{\smash{{\SetFigFont{6}{7.2}{\rmdefault}{\mddefault}{\updefault}{\color[rgb]{0,0,0}$V_2$}%
}}}}
\put(1760,-625){\makebox(0,0)[lb]{\smash{{\SetFigFont{6}{7.2}{\rmdefault}{\mddefault}{\updefault}{\color[rgb]{0,0,0}$V_3$}%
}}}}
\put(1612,-2306){\makebox(0,0)[lb]{\smash{{\SetFigFont{6}{7.2}{\rmdefault}{\mddefault}{\updefault}{\color[rgb]{0,0,0}$V_3$}%
}}}}
\put(3270,-183){\makebox(0,0)[lb]{\smash{{\SetFigFont{6}{7.2}{\rmdefault}{\mddefault}{\updefault}{\color[rgb]{0,0,0}$\Omega_1$}%
}}}}
\put(2439,-146){\makebox(0,0)[lb]{\smash{{\SetFigFont{6}{7.2}{\rmdefault}{\mddefault}{\updefault}{\color[rgb]{0,0,0}$V_1$}%
}}}}
\put(3640,445){\makebox(0,0)[lb]{\smash{{\SetFigFont{6}{7.2}{\rmdefault}{\mddefault}{\updefault}{\color[rgb]{0,0,0}$V_2$}%
}}}}
\put(3722,-686){\makebox(0,0)[lb]{\smash{{\SetFigFont{6}{7.2}{\rmdefault}{\mddefault}{\updefault}{\color[rgb]{0,0,0}$V_3$}%
}}}}
\put(2425,-1798){\makebox(0,0)[lb]{\smash{{\SetFigFont{6}{7.2}{\rmdefault}{\mddefault}{\updefault}{\color[rgb]{0,0,0}$V_1$}%
}}}}
\put(3856,-1213){\makebox(0,0)[lb]{\smash{{\SetFigFont{6}{7.2}{\rmdefault}{\mddefault}{\updefault}{\color[rgb]{0,0,0}$V_2$}%
}}}}
\put(3722,-2342){\makebox(0,0)[lb]{\smash{{\SetFigFont{6}{7.2}{\rmdefault}{\mddefault}{\updefault}{\color[rgb]{0,0,0}$V_3$}%
}}}}
\put(1089,-278){\makebox(0,0)[lb]{\smash{{\SetFigFont{6}{7.2}{\rmdefault}{\mddefault}{\updefault}{\color[rgb]{0,0,0}$\Omega_2$}%
}}}}
\put(1081,-1911){\makebox(0,0)[lb]{\smash{{\SetFigFont{6}{7.2}{\rmdefault}{\mddefault}{\updefault}{\color[rgb]{0,0,0}$\Omega_1$}%
}}}}
\put(2842,-1899){\makebox(0,0)[lb]{\smash{{\SetFigFont{6}{7.2}{\rmdefault}{\mddefault}{\updefault}{\color[rgb]{0,0,0}$\Omega_1$}%
}}}}
\put(3265,-1774){\makebox(0,0)[lb]{\smash{{\SetFigFont{6}{7.2}{\rmdefault}{\mddefault}{\updefault}{\color[rgb]{0,0,0}$\Omega_3$}%
}}}}
\put(3538,-1974){\makebox(0,0)[lb]{\smash{{\SetFigFont{6}{7.2}{\rmdefault}{\mddefault}{\updefault}{\color[rgb]{0,0,0}$\Omega_2$}%
}}}}
\end{picture}%

%% file: extension.pdf_t
\begin{picture}(0,0)%
\includegraphics{extension.pdf}%
\end{picture}%
\setlength{\unitlength}{4144sp}%
\begingroup\makeatletter\ifx\SetFigFont\undefined%
\gdef\SetFigFont#1#2#3#4#5{%
  \reset@font\fontsize{#1}{#2pt}%
  \fontfamily{#3}\fontseries{#4}\fontshape{#5}%
  \selectfont}%
\fi\endgroup%
\begin{picture}(3540,1098)(121,-406)
\put(3646,-331){\makebox(0,0)[lb]{\smash{{\SetFigFont{8}{9.6}{\rmdefault}{\mddefault}{\updefault}{\color[rgb]{0,0,0}$p$}%
}}}}
\put(3016,-222){\makebox(0,0)[lb]{\smash{{\SetFigFont{6}{7.2}{\rmdefault}{\mddefault}{\updefault}{\color[rgb]{0,0,0}$\omega$}%
}}}}
\put(2071,434){\makebox(0,0)[lb]{\smash{{\SetFigFont{6}{7.2}{\rmdefault}{\mddefault}{\updefault}{\color[rgb]{0,0,0}$\Pcsmax$}%
}}}}
\put(1711,-331){\makebox(0,0)[lb]{\smash{{\SetFigFont{8}{9.6}{\rmdefault}{\mddefault}{\updefault}{\color[rgb]{0,0,0}$p$}%
}}}}
\put(136,119){\makebox(0,0)[lb]{\smash{{\SetFigFont{6}{7.2}{\rmdefault}{\mddefault}{\updefault}{\color[rgb]{0,0,0}$\Pcsmax$}%
}}}}
\put(1081,-222){\makebox(0,0)[lb]{\smash{{\SetFigFont{6}{7.2}{\rmdefault}{\mddefault}{\updefault}{\color[rgb]{0,0,0}$\omega$}%
}}}}
\put(556,-364){\makebox(0,0)[lb]{\smash{{\SetFigFont{6}{7.2}{\rmdefault}{\mddefault}{\updefault}{\color[rgb]{0,0,0}$\omega\mathord{-}\Pcsmax$}%
}}}}
\put(2277,344){\makebox(0,0)[lb]{\smash{{\SetFigFont{6}{7.2}{\rmdefault}{\mddefault}{\updefault}{\color[rgb]{0,0,0}$\omega$}%
}}}}
\put(342,344){\makebox(0,0)[lb]{\smash{{\SetFigFont{6}{7.2}{\rmdefault}{\mddefault}{\updefault}{\color[rgb]{0,0,0}$\omega$}%
}}}}
\put(485,569){\makebox(0,0)[lb]{\smash{{\SetFigFont{8}{9.6}{\rmdefault}{\mddefault}{\updefault}{\color[rgb]{0,0,0}$\Pc(p)$}%
}}}}
\put(2420,569){\makebox(0,0)[lb]{\smash{{\SetFigFont{8}{9.6}{\rmdefault}{\mddefault}{\updefault}{\color[rgb]{0,0,0}$\Pc(p)$}%
}}}}
\end{picture}%